\documentclass[11pt,a4paper]{article}
\usepackage{jcappub}
\usepackage{float}
\usepackage{rotating}
\usepackage{graphicx}
\usepackage{epsfig}
\usepackage{amsmath}
\usepackage{amssymb}
\usepackage{mathrsfs}  
\usepackage{subfigure}
\usepackage{time}
\usepackage{relsize, comment}
\usepackage{xcolor}
\usepackage[utf8]{inputenc}

\newcommand{\snote}[1]{\textcolor{red}{{\bf [Sh: #1]}}}

\hypersetup{ linktoc=all,
    colorlinks=true, linkcolor={blue},  
       citecolor={red}, urlcolor={vividviolet}
}
\definecolor{rosy}{RGB}{230,235,252}
\definecolor{myframetitle}{RGB}{90,89,170}
\definecolor{myblocktitle}{RGB}{140,185,249}
\definecolor{mytitle}{RGB}{10,80,26}

\definecolor{darkgreen}{RGB}{27,130,45}
\definecolor{darkblue}{rgb}{0,0,0.3}
\definecolor{darkred}{rgb}{0.7,0,0}

\definecolor{light gray}{RGB}{220,220,220}
\definecolor{dark purple}{RGB}{108,0,217}
\definecolor{pink}{RGB}{190,20,100}
\definecolor{orang}{RGB}{193,63,0}
\definecolor{green}{RGB}{11,98,17}
\definecolor{darkpink}{RGB}{153,0,76}
\definecolor{bluegreen}{RGB}{0,102,102}
\definecolor{greenlagan}{RGB}{0,102,0}
\definecolor{redgreen}{RGB}{102,102,0}
\definecolor{Redgreen}{RGB}{153,76,0}
\definecolor{vividviolet}{rgb}{0.62, 0.0, 1.0}
\definecolor{amaranth}{rgb}{0.9, 0.17, 0.31}
\definecolor{palatinateblue}{rgb}{0.15, 0.23, 0.89}
\definecolor{brightpink}{rgb}{1.0, 0.0, 0.5}
\definecolor{cornflowerblue}{rgb}{0.39, 0.58, 0.93}
\definecolor{deepcarminepink}{rgb}{0.94, 0.19, 0.22}
\definecolor{radicalred}{rgb}{1.0, 0.21, 0.37}

\title{Big Bang in Dipole Cosmology}

\author[a]{A. Allahyari,}\emailAdd{alireza.al@ipm.ir}
\author[b]{E. Ebrahimian,}\emailAdd{ehsanebrahimianarejan@gmail.com}
\author[c]{R. Mondol,}\emailAdd{ranjinim@iisc.ac.in}
\author[d,b]{M.M. Sheikh-Jabbari}\emailAdd{jabbari@theory.ipm.ac.ir}

\affiliation[a]{Department of Astronomy and High Energy Physics, Faculty of Physics, Kharazmi University, Tehran 15719-14911, Iran}

\affiliation[b]{The Abdus Salam ICTP, Strada Costiera 11, I-34014 Trieste, Italy}

\affiliation[c]{Center for High Energy Physics,
Indian Institute of Science, Bangalore 560012, India
}

\affiliation[d]{School of Physics, Institute for Research in Fundamental
Sciences (IPM),\\ P. O. Box 19395-5531, Tehran, Iran}

\abstract{We continue the study of dipole cosmology framework put forward in \cite{Krishnan:2022qbv}, a beyond FLRW setting that has a preferred direction in the metric which may be associated with a cosmological tilt, a cosmic dipole. In this setup the shear and the tilt can be positive or negative given the dipole direction. We thoroughly analyze evolution of the universe in this setting, particularly focusing on the behaviour near the Big Bang (BB). We first analyze a single fluid model with a generic constant equation of state $w$. 
While details of the behavior near the BB depends on $w$ and the other initial conditions, we find that when the shear is negative we have a shear dominated BB singularity,  whereas for a positive shear we have a much milder singularity, the whimper singularity \cite{Ellis:1974ug}, at which the tilt blows up while curvature invariants remain finite. We then consider dipole $\Lambda$CDM model  which besides the shear has two tilt parameters, one for radiation and one for the pressureless matter. For positive (negative) shear we again find whimper (curvature) singularity near the BB. Moreover, when the tilt parameters have opposite signs, the shear can change sign from negative to positive in the course of evolution of the Universe. We show that the relative tilt of the radiation and the matter generically remains sizable at late times.}

\begin{document}
\maketitle
\flushbottom

\section{Introduction}

The assumption that the Universe is homogeneous and isotropic at cosmological scales,  usually  called the cosmological principle, has dominated  cosmological studies over the last century. Such cosmologies at the background level are described by Friedmann–Lemaître–Robertson –Walker (FLRW) models. Before the inflation scenario became  a staple of the standard model of cosmology, homogeneity and isotropy of the Universe were viewed as an improbable situation \cite{Collins:1972tf}. Within inflationary settings, however, thanks to Wald's cosmic no-hair theorem \cite{Wald:1983ky}, isotropy seems to be a generic fixed point of evolution in the late Universe, if we have a positive cosmological constant and assume homogeneity.\footnote{Inflationary models generically do not obey  assumptions of Wald's theorem and hence they are not subject to the theorem. Nonetheless, improving on Wald's theorem, it has been shown that inflationary models generically suppress isotropy, while not necessarily washing it out \cite{Maleknejad:2012as}. Moreover, it has been argued that an anisotropy signature, a dipole component, can remain in the superhorizon modes generated during inflation \cite{Turner:1991dn}. Such mode may have a chance to reenter the horizon afterward and induce a dipole in the galaxy number-counts \cite{Domenech:2022mvt}. }

In the era that  a good wealth of cosmological data is or will be available, all cosmological assumptions including the cosmological principle  should be tested against the observational data. While it seems to be compatible to a good degree with the existing data, accumulation of different datasets at various redshifts and scales has led to the appearance of anomalies in homogeneity or isotropy, see \cite{Aluri:2022hzs} for a comprehensive recent review.  In particular, there are several observations pointing to the existence of a (non-kinematical) cosmic dipole \cite{Colin:2019opb, Singal:2021crs, Secrest:2020has, Krishnan:2021jmh, McConville:2023xav, Yeung:2022smn, Watkins:2023rll, Bengaly:2018ykb, Bengaly:2019zhr}.\footnote{We note that there are many notable reports of finding ``bulk flows'' within astrophysical data. The bulk flows have been observed at distances of the size and distance of superclusters (few 100 Mpc) e.g. see \cite{Tully:2022rbj, Whitford:2023oww, Hoffman:2015waa} or within other cosmological datasets like the cosmic microwave background \cite{Lavaux:2012jb} or supernovae \cite{Dai:2011xm}. Whether the bulk flows are of cosmological origin and/or is related to the background cosmological dipoles, like those formulated within our setting here, is not clear with the data currently available. } These hints for a cosmic dipole are not conclusive yet and one need more and better quality data. In waiting for such data, it is prudent to explore construct and study models that can accommodate a cosmic dipole. One class of these models which is the main focus of this paper, is based on the  the dipole cosmological principle  introduced in \cite{Krishnan:2022qbv} and further explored in \cite{Krishnan:2022uar, Ebrahimian:2023svi}. It assumes homogeneity and axisymmetry and existence of locally boosted frames that exhibit local rotation symmetry. 

Dipole cosmology is formulated within the tilted cosmology setting of King and Ellis \cite{King:1972td}. In the tilted cosmology setup, one starts with a homogeneous background, i.e. a Bianchi cosmology, see \cite{Ellis:1998ct, EMM-2012} for thorough discussions and reviews. Given a Bianchi cosmology metric, one can ask if the energy momentum tensor has a nonzero momentum flow through  constant time slices (where we find homogeneous 3 dimensional spaces). That is, if we denote the time coordinate in the Bianchi metric by $t$ and the energy momentum tensor by $T^\mu{}_\nu$, the question is whether $P_\nu=T^t{}_\nu$ is a vector along the $t$ direction or not. If we assume dominant energy condition (see appendix \ref{appendix}), which is usually assumed in cosmology, { $P_\nu$ is a (future oriented) timelike vector  and its projection on 3 dimensional constant $t$ spatial slices defines the tilt vector}.\footnote{In \cite{Ebrahimian:2023svi} and also in this work, while $P_\mu$ is still  future oriented, we allow its spatial part to be in the same or opposite direction as the dipole, i.e. the tilt or shear can have either signs.} There is hence  a boosted frame with rapidity $\beta$, the tilt, along this spatial vector in which $P_\nu$ is along the time direction. Existence of such tilts can be traced to the cosmic dipole reported in the observations \cite{Ebrahimian-in-preparation}.\footnote{In the tilted cosmology setup the tilt parameter $\beta$ is in general a function of cosmic time and background is anisotropic. There are class of non-comoving cosmologies where they assume background metric to be FLRW type but the cosmic fluid is not in a comoving frame \cite{Cembranos:2019plq, Ganguly:2017qff, Najera:2020smt}.}
Dynamics of simple tilted cosmological models within the King-Ellis setup \cite{King:1972td, Ellis:1998ct} has been studied e.g. see \cite{Coley:2006nw, Dagwal:2018apx, Coley:2004jm, Coley:2004jj, Barrow:2003fc, Goliath:1998na, Hewitt:1992sk, Goliath:2000ag, Bradley:2005kk} and \cite{Santiago:2022yjt, Tsagas:2021tqa, Tsagas:2015mua}. 

In this work we continue our analysis within  the dipole cosmology setting and focus on the early Universe and evolution of such models near the Big Bang (BB). In particular, we explore the singularity structure where the effective overall scale factor of these cosmologies approaches zero.
Such analyses, especially within the Bianchi models predate tilted models and inflation idea. Belinsky-Khalatnikov-Lifshitz (BKL) studied  Kasner type models in search for a possible resolution for the horizon problem \cite{Belinsky:1970ew}. Similar analysis was repeated for closed Universes in the so-called mixmaster Universe \cite{Misner:1969hg}. The analysis of singularities in Bianchi models is not limited to the BB and may also be relevant to future (big crunch) singularities. See \cite{Belinsky:1970ew, Ellis:1974ug, Ellis:1977pj, Madsen:1986mi, Ellis:2006ba, Joshi:2013xoa, Lubbe:2023aql, Erickson:2003zm} and specially \cite{Collins-1974,COLLINS:1979, Barrow:2002is} for some related work.

For single fluid dipole cosmologies  in the limit where overall scale factor vanishes we find two generic  behaviours: either the tilt grows and becomes large while the matter density remains finite (known as whimper singularity \cite{Ellis:1974ug}) or tilt becomes very small while the matter density blows up, but the geometry is extremely anisotropic and the shear is very close to its maximal value. Our results here improves and extends those in the earlier works \cite{Collins-1974, COLLINS:1979, Barrow:2002is} for the dipole cosmologies. {Among these, \cite{COLLINS:1979} discusses singularity structure in Bianchi cosmologies, in particular Bianchi V tilted cosmology, which is also the main setup of dipole cosmologies. In \cite{COLLINS:1979} the focus is on the single fluid models with equation of state $0\leq w\leq 1$ ($\rho \geq p>0$, where $\rho, p$ are energy density and pressure of the fluid). In our analysis here we expand and generalize these analysis in some different ways.  Our analysis is more quantitative than Collins and Ellis work \cite{COLLINS:1979}, we supplement analytic treatments with numeric analysis; we consider equation of state to be in $-1\leq w\leq 1$ and allow for negative as well as positive tilt and also provide a more detailed analysis of curvature and its singular behaviour. These yield interesting features and singularity behaviour not discussed in \cite{COLLINS:1979}. Moreover, we discuss multi-fluid cases, in particular, the case of dipole $\Lambda$CDM model \cite{Ebrahimian:2023svi}.}

Dipole $\Lambda$CDM \cite{Ebrahimian:2023svi} is a specific case with three fluids, radiation, pressureless matter and dark energy. Tilted cosmologies with two fluids have been discussed in \cite{Goliath:2000ag, Coley:2004jj}, mainly focusing on the late time behaviour in such models. In our model the dark energy is modeled by the 
cosmological constant $\Lambda$, which is tilt inert \cite{Krishnan:2022qbv}, therefore, dipole $\Lambda$CDM constitutes an example of a tilted  two-fluid cosmology. We exhaust all possibilities for the near BB behaviour in dipole $\Lambda$CDM model and show that we either have a whimper singularity or a curvature singularity. We present our numerical analyses plots with parameters compatible with  Planck collaboration values for $\Lambda$CDM parameters \cite{Planck-2018}.

The rest of this paper is organized as follows. In section \ref{sec:2-basic-setup} we briefly introduce the basics of dipole cosmology and discuss some of its general features. In section \ref{sec3:near-BB-general-w} we focus on the single fluid case and behavior of the system near the BB. We make an exhaustive study for cosmic fluids respecting dominant energy condition. In section \ref{sec4:Dipole-LCDM} we study the dipole $\Lambda$CDM \cite{Ebrahimian:2023svi} near the BB as well as the evolution of the model till today and in near future. We close in section \ref{sec5:closing} by concluding remarks and outlook. To make the paper more self-contained, in appendix \ref{appendix}  we have briefly reviewed energy conditions for the tilted fluids.

\section{Dipole Cosmology, the basic setup and some general features}\label{sec:2-basic-setup}

We start with a review of the basic equations for dipole cosmologies. This section is not just a review of the original papers \cite{Krishnan:2022qbv, Krishnan:2022uar, Ebrahimian:2023svi}, it expands and extends the analysis there especially regarding the (curvature) singularity structure as well as adding new general discussions and insights. Dipole cosmology has two basic ingredients, metric and energy momentum tensor.
\paragraph{Metric.} The anisotropic background metric of the Bianchi type V has the form,
\begin{equation}\label{DipoleMetric}
    ds^{2} = -dt^{2} + a^{2}(t)\left[ e^{4b(t)} dz^{2} +  e^{-2b(t)-2A_{0} z}\big(dx^{2}+dy^{2}\big) \right],
\end{equation}
where $a(t)$ is the overall scale factor, $b(t)$ captures the anisotropy and $A_0\neq 0$ is a constant with dimension of inverse length. {As it is well known Bianchi V models describe an open Universe and $A_0$ is parametrizing the spatial curvature.} $A_0$ can be set to 1 by a choice of units, but we will keep it for later convenience and assume it takes positive values.
\paragraph{Tilted  energy momentum tensor.} Let $u^\mu$ denote the velocity of the fluid in the comoving  frame in $(t,z, x,y)$ coordinates.  For a tilted fluid which moves with rapidity $\beta$  we have \cite{King:1972td},
\begin{equation}\label{u-mu}
    u^\mu= (cosh\beta,  (a e^{2b})^{-1}\sinh\beta, 0, 0),\quad u_\mu= (-cosh\beta, (a e^{2b})\sinh\beta , 0, 0),\quad u^2=-1.
\end{equation}
The energy momentum tensor for the tilted fluid is then,
\begin{equation}\label{T-uu}
    T^\mu{}_\nu= \rho u^\mu u_\nu + p (\delta^\mu{}_\nu+u^\mu u_\nu),
\end{equation}
more explicitly,
\begin{equation}\label{T-tilted}
\begin{split}
    &T^{\mu}{}_{\nu} = T_{\text{iso}}{}^{\mu}{}_{\nu} + T_{\text{tilt}}{}^{\mu}{}_{\nu},\qquad 
T_{\text{iso}}{}^{\mu}{}_{\nu}=\text{diag}(-\rho, p,p,p)\\
&T_{\text{tilt}}{}^{\mu}{}_{\nu}=(\rho+p)\sinh\beta\left(\begin{array}{cccc}
 -\sinh\beta &  ae^{2b}  \cosh \beta & 0 & 0 \\
-\cosh \beta/(ae^{2b}) &  \sinh\beta  & 0 & 0 \\
 0 & 0 & 0 & 0 \\
 0 & 0 & 0 & 0\\
\end{array}
\right)
\end{split}
\end{equation}
where $T_{\text{iso}}{}^{\mu}{}_{\nu}$ is the energy momentum tensor of a typical isotropic perfect fluid and $\rho$ and $p$ denote its  energy density  and pressure. Recalling the form of energy momentum tensor \eqref{T-uu}, one can readily see that
\begin{equation}\label{T-trace-T2}
    T:=T^\mu{}_\mu=-\rho+ 3p,\qquad T^\mu{}_\nu T^\nu{}_\mu= \rho^2+3p^2,
\end{equation}
and in general any scalar built from $T^\mu{}_\nu$ is $\beta$ independent. This follows from the definition of the tilt \cite{King:1972td} (briefly reviewed in the introduction) and the fact that there exists a frame comoving with the fluid where the energy momentum tensor $T^\mu{}_\nu$ is $\beta$ independent and diagonal.  

Consider a multi-fluid case consisting of $N$ \textit{non-interacting} fluid components. To specify the system besides energy density and pressure for each component, $\rho_i, p_i, i=1,\cdots, N$, we need  to associate a tilt to each component $\beta_i$. While in general the tilts for each component need not be in the same direction, to respect the dipole cosmological principle we take the flows all to be in the same direction $z$ while their value can be different. See \cite{Ebrahimian:2023svi} for more discussions. For this case, therefore, 
\begin{equation}\label{T-sum}
T^{\mu}{}_{\nu} = \sum_{i=1}^N (T^{\mu}{}_{\nu}){}_i, \qquad \nabla_\mu(T^{\mu}{}_{\nu}){}_i=0\quad \forall i,
\end{equation} where $(T^{\mu}{}_{\nu}){}_i$ has the same expression as  \eqref{T-tilted} with $\rho=\rho_i,\ p=p_i,\ \beta=\beta_i$. 

\subsection{Basic analysis of the dipole models} One may insert the above into Einstein equations (in our units $8\pi G=1$)
\begin{equation}\label{Ein-Eq}
    R_{\mu\nu}-\frac12 R g_{\mu\nu}=\Lambda g_{\mu\nu}+ T_{\mu\nu},
\end{equation}
and continuity equations \eqref{T-sum} to obtain
\begin{subequations}\label{continuty-eq}
\begin{align}
\dot{\rho}_i+3H(\rho_i+p_i)&=-(\rho_i+p_i)\tanh\beta_i(\dot{\beta}_i-\frac{2A_0}{a} e^{-2b}) \label{Con-rho-i}, \\
\dot{p}_i+H(\rho_i+p_i)&= -(\rho_i+p_i)\left( \frac23\sigma+\dot{\beta}_i\coth{\beta}_i\right). \label{Con-p-i}
\end{align}
\end{subequations}
\begin{subequations}\label{Einstein-eq}
\begin{align}
&H^2-\frac19\sigma^2-\frac{A_0^2}{a^2} e^{-4b}=\frac{\Lambda}{3}+\sum_i\frac{\rho_i}{3}+\frac13 (\rho_i+p_i)\sinh^2\beta_i ,
\label{EoM-H-sigma-sum}\\
&\sigma =\frac{1}{4A_0} a e^{2b}\ \sum_i(\rho_i+p_i)\sinh2\beta_i,
\label{EoM-sigma-sum}
\end{align}
\end{subequations}
where as usual 
\begin{equation}
    H:= \frac{\dot{a}}{a},\qquad \sigma:= 3\dot{b}. \label{Definition-H-sigma}
\end{equation}
With our conventions $A_0>0$ while $\sigma, \beta_i$ can be positive or negative. Assuming null energy condition (NEC) $\rho_i+p_i\geq 0$, \eqref{EoM-sigma-sum} tells us that sign of $\sigma$ and sign of $\beta_i$ are related: if all $\beta_i$ are positive (or negative) $\sigma$ is positive (or negative); if $\beta_i$ have different signs, then $\sigma$ may have different signs and interestingly, it can change sign in the course of evolution, see \cite{Ebrahimian:2023svi} and also section \ref{sec4:Dipole-LCDM}. 
Einstein equations \eqref{Ein-Eq} also yield
\begin{subequations}\label{Hdot-sigmadot}
\begin{align}
&\dot{H}+H^2=\frac{\Lambda}{3}-\frac16\ \sum_i [{(\rho_i+3p_i)}+2(\rho_i+p_i)\sinh^2\beta_i]-\frac29 \sigma^2
\label{EoM-acceleration-0}\\
&\dot{\sigma}+3H\sigma =\sum_i(\rho_i+p_i )\sinh^2\beta_i. \label{EoM-H-sigma-sigma-dot}
\end{align}\end{subequations}
The  above equations are not independent from \eqref{continuty-eq}, \eqref{Einstein-eq}  and we have added them because they are helpful in our near Big Bang analysis below. Moreover, \eqref{EoM-H-sigma-sigma-dot} reveals an interesting feature. Assuming strict null energy condition $\rho_i+p_i> 0$, $\dot{\sigma}+3H\sigma>0$. This means that if $\sigma<0$ then $\dot{\sigma}>0$ and conversely, if $\dot{\sigma}<0$ then $\sigma>0$. This in particular implies $\sigma$ can only change sign from negative to positive (as Universe evolves forward in time) and not from positive to negative \cite{Ebrahimian:2023svi}. 

For a positive $\Lambda$, which we assume, it is useful to define energy density ratios as usual:
\begin{equation}\label{Omega-def}
    \Omega_\Lambda=\frac{\Lambda}{3H^2},\qquad \Omega_k:=\frac{A_0^2}{H^2a^2 e^{4b}},\qquad \Omega_\sigma:=\frac{\sigma^2}{9H^2},\qquad \Omega_i:=\frac{\rho_i+(\rho_i+p_i)\sinh^2\beta_i}{3H^2}.
\end{equation}
Weak energy condition (WEC) $\rho_i\geq 0$ which we assume and \eqref{EoM-H-sigma-sum} yield,
\begin{equation}\label{Omega-sum-rule}
    \Omega_\Lambda+\Omega_k+\Omega_\sigma+\sum_i \Omega_i=1,\qquad 0\leq \Omega_X\leq 1,\ \ \forall X.
\end{equation}
For later use, it will be handy to introduce,
\begin{equation}\label{kappa-K-S}
    \kappa:=\sqrt{\Omega_k}=\frac{A_0}{Ha e^{2b}},\qquad  \sigma:=3H\mathfrak{S}K, \qquad  \qquad 0\leq \kappa, K\leq 1, \quad \mathfrak{S}=\pm 1.
\end{equation}
That is, $\mathfrak{S}$ is defining sign of $\sigma$, for positive (or negative) $\sigma$,  $\mathfrak{S}$ is $+1$ (or $-1$) and $\Omega_\sigma=K^2$. Note that $\mathfrak{S}$ is in principle a function of time and can change sign in the course of evolution. However, as discussed above, $\mathfrak{S}$ can only evolve from negative to positive (and not the reverse). {From  \eqref{Definition-H-sigma} and \eqref{kappa-K-S} we learn that}
\begin{equation}
    b=\int_0^a \frac{\mathfrak{S}K}{a}\ da ,
\end{equation}
where we have chosen the integration constant such that $b(a\to 0)=0$.

\subsection{Curvature analysis} 
Singularities in spacetimes has been classified in \cite{Ellis:1977pj, Clarke:1977pi} and in most cases relevant to cosmology these are curvature singularities, i.e. there is a locus in spacetime where some of curvature invariants blow up.  
Riemann tensor appears in the geodesic deviation equation. The Riemann tensor components can be expressed in tetrad basis as
\begin{equation}
 \Tilde{R}_{abcd}=R_{\mu\nu\alpha\beta}e^\mu_a e^\nu_b e^\alpha_c e^\beta_d ,  
\end{equation}
where $a,b,c,d$ are tetrad indices and $g_{\mu\nu}e^\mu_a e^\nu_b=\delta_{ab}$.  The non zero components are
\begin{subequations}\label{Riemann-frame}
\begin{align}
  &\Tilde{R}_{tyty}=\Tilde{R}_{txtx}= -H^2 \left(\frac{aH'}{H}-\frac{a\sigma'}{3H}
  + (1-\frac{\sigma}{3H})^2\right),\\
 & \Tilde{R}_{tztz}=
 -H^2 \left(\frac{aH'}{H}+\frac{2a\sigma'}{3H}
 + (1+\frac{2\sigma}{3H})^2\right),\\
 & \Tilde{R}_{tyyz}=\Tilde{R}_{txxz}=\sigma \kappa H,\\ 
 & \Tilde{R}_{yzyz}=\Tilde{R}_{zxzx}= H^2\left(1-\kappa^2+\frac{\sigma}{3H}-\frac{2\sigma^2}{9H^2}\right),\\
 & \Tilde{R}_{xyxy}= H^2\left((1-\frac{\sigma}{3H})^2-\kappa^2\right). 
 \end{align}
\end{subequations}

Riemann curvature may be decomposed in Ricci tensor and Weyl curvature. The former is specified through Einstein's equations and a singularity in Ricci means a singularity in the energy-momentum tensor of the matter filling the spacetime. The Weyl curvature $W_{\mu\nu\alpha\gamma}$  may be divided into its symmetric traceless electric and magnetic parts, respectively defined as
\begin{equation}
\mathcal{E}_{\mu\nu}:=W_{\mu\alpha\nu\gamma} v^\alpha v^\gamma, \qquad \mathcal{M}_{\mu\nu}:=\frac12 \epsilon_{\mu\alpha\rho\gamma}W^{\rho\gamma}{}_{\nu\delta} v^\alpha v^\delta ,
\end{equation}
where $\epsilon_{\mu\alpha\rho\gamma}$ is the Levi-Civita totally antisymmetric tensor and $v^\mu$ is a unit norm timelike vector $v^2=-1$. We can take $v^\mu=(1,0,0,0)$, as the velocity vector in metric comoving frame or can be chosen as fluid velocity vector \eqref{u-mu}; here we choose the former. It is well known that the magnetic part of the Weyl tensor vanishes due to homogeneity and the electric part also vanishes for homogeneous and isotropic FLRW cosmologies, e.g. see \cite{Barrow:2002is, Gurses:2020kpv}. In our case, dealing with a homogeneous cosmology, magnetic part vanishes, but the electric part is non-vanishing, basically measuring the cosmic shear and its time variation. For the metric ansatz \eqref{DipoleMetric} its non-vanishing components are,
\begin{equation}\label{Electric-Weyl}
    \mathcal{E}^x_{\;x}=\mathcal{E}^y_{\;y}=-\frac12\mathcal{E}^z_{\;z}= \frac{\sigma H}{6} \left(1+ \frac{2\sigma}{3H}+\frac{a\sigma'}{\sigma}\right).
\end{equation}

For later use we also present the Kretschmann scalar,
\begin{equation}\label{Kretschmann}
\begin{split}
\mathcal{K}=R^{\mu\nu\alpha\gamma}R_{\mu\nu\alpha\gamma}=12H^4\biggl[&(1-\kappa^2)^2+(3K^2+1)^2+2K^2(3-5\kappa^2+2K\mathfrak{S})\\
+&2K^2 (\frac{a\sigma'}{\sigma})(\frac{a\sigma'}{\sigma}+4+2K\mathfrak{S})+(\frac{aH'}{H})(\frac{aH'}{H}+2+4K^2)\biggr].
\end{split}    
\end{equation}

\subsection{Single fluid case}

Let us start  with a universe with a single fluid with a constant EoS case, $p=w \rho$, here we assume dominant energy condition (DEC), i.e. $\rho>0$ and $ -1 <w\leq 1$. See appendix \ref{appendix} for a short review on energy conditions.  For this case {\eqref{continuty-eq} and  \eqref{Einstein-eq} or their integrated form yield \cite{Krishnan:2022qbv, Krishnan:2022uar}}
\begin{subequations}\label{const-w-dipole-1}
\begin{align}
&H^2=\frac{\Lambda}{3}+\frac{A_0^2}{a^2}e^{-4b}+\frac{\rho}{3}\left[1+(w+1)\sinh^2\beta\right]+\frac19 \sigma^2\label{hubble}
\\
&{\beta}' \big(\coth\beta-w \tanh \beta\big)=\frac{3w-1}{a}-\frac2{3a}\frac{\sigma}{H}-\frac{2w A_0}{a^2 H} e^{-2b}\tanh\beta \label{beta-growth-w-const}
\\
&\rho^{\frac{w}{1+w}} a e^{2b} \sinh\beta =  C = const.
\label{const-w-dipole-rho-X-beta}\\
&\sigma=\frac{C (1+w)}{2A_0} \rho^{\frac{1}{1+w}}\cosh\beta=\frac{(1+w)}{4A_0} a e^{2b}\ \rho\ \sinh2\beta \label{sigma-w-dipole}\\
&  \dot{H}+H^2=\frac{\Lambda}{3}-\frac{\rho}{6}(1+3w)-\frac{\rho}{3}(1+w)\sinh^2\beta-\frac29 \sigma^2.
\label{EoM-acceleration} 
\end{align}
\end{subequations}
{Eq.~\eqref{const-w-dipole-rho-X-beta} is obtained by integrating one of the continuity equations where $C$ is the integration constant and prime denotes derivative w.r.t. $a$, $X':=\frac{dX}{da}$; one may use $a$ as the cosmic clock.} We also note that \eqref{EoM-H-sigma-sigma-dot} and \eqref{EoM-acceleration-0} take the form
\begin{subequations}\label{const-w-dipole-2}
\begin{align}
{\sigma}'+\frac{3}{a}\sigma &=\frac{\rho}{aH} (1+w)\sinh^2\beta,\qquad \sigma= 3a H b' \label{EoM-H-sigma-sigma-dot-a}\\
H^2(1+\frac{aH'}{ H})&=\frac{\Lambda}{3}-\frac{\rho}{6}(1+3w)-\frac{\rho}{3}(1+w)\sinh^2\beta-\frac29 \sigma^2.
\label{EoM-acceleration-1}
\end{align}\end{subequations}

\subsection{General power laws}\label{sec:general-power-law}

It is useful to explore the log-derivative of our physical parameters w.r.t. $a$, explicitly,
\begin{subequations}\label{log-log-exponents}
\begin{align}
r &:=\frac{d\ln\rho}{d\ln a}=-\frac{1+w}{1-w\tanh^2\beta}\left(3-2\kappa\tanh\beta-(1+2K\mathfrak{S})\tanh^2\beta \right), \label{r-def}\\
s&:=3+\frac{d\ln \sigma}{d\ln a}=2\sqrt{\Omega_k}\tanh{\beta}=2\kappa \tanh\beta ,\label{s-def}\\
h&:=\frac{d\ln H}{d\ln a}
=-\kappa^2-3K^2-2\Omega_w -\frac{3w-1}{2[1+(1+w)\sinh^2\beta]}\Omega_w, \label{h-def}\\ 
f&:=\frac{d\ln \sinh\beta}{d\ln a}=\frac{1}{1-w\tanh^2\beta}\left(3w-1-2\mathfrak{S}K- 2w \kappa \tanh\beta \right),\label{f-def}\\
F&:=\frac{d\ln \cosh\beta}{d\ln a}=f\tanh^2\beta=-3-\frac{r}{1+w}+2\kappa \tanh\beta, \label{F-def}
\end{align}
\end{subequations}
where in the above we used  \eqref{const-w-dipole-1} and \eqref{const-w-dipole-2}. Equations \eqref{r-def}, \eqref{f-def} and \eqref{F-def} are not independent. Note that the above equations {are true at all times (any $a$) and } may be viewed as replacements for the dynamical equations and {therefore, the exponents} $r,s,h,f,F$ are also in general functions of $a$. Moreover, from \eqref{const-w-dipole-1} and definitions \eqref{Omega-def}, we learn
\begin{equation}\label{Omega-w-Kk}
    \Omega_w=2K\kappa (\frac{w}{1+w}|\tanh\beta|+\frac{1}{1+w}\frac1{|\tanh\beta|}) > 2K\kappa.
\end{equation}

The above equations have the following implications.
\begin{enumerate}
\item \textbf{Bound on $\boldsymbol{\kappa+ K}$:} Since $\Omega_m+\Omega_\sigma+\Omega_k<1, \Omega_\sigma=K^2, \Omega_k=\kappa^2$, \eqref{Omega-w-Kk} yields 
\begin{equation}\label{k-K-ineq}
    (\kappa+K)^2 <1. 
\end{equation}
{While $(\kappa+K)^2$ can be arbitrarily close to $1$, it should always remain less than 1. }
\item \textbf{Bound on $\boldsymbol{s}$:} Similarly \eqref{s-def} implies that $-2\leq s\leq 2$ or $1\leq d\ln\sigma/d\ln a \leq 5$ for the whole course of the evolution of the Universe. 

\item \textbf{Discussion on the sign $\mathfrak{S}$:}  The{factor inside parenthesis} in the RHS of \eqref{r-def} {can} be written as 
    \begin{equation}\label{r-S-sign}
        3-2\kappa\tanh\beta-(1+2K\mathfrak{S})\tanh^2\beta= 1-\tanh^2\beta+2[1-\mathfrak{S}(K\tanh^2\beta+\kappa |\tanh\beta|)] ,
    \end{equation}
where we used that $\beta$ has the same sign as $\mathfrak{S}$ and that $K,\kappa, |\tanh\beta|\leq 1$. 
{The quantity above is positive for $\mathfrak{S} = -1$ at any time. This implies for $\mathfrak{S} = -1$, always $r < 0$.}  For this case, however, $f$ can have either sign depending on $w$. {In the next subsection, we will explore the behaviour of these exponents near the singularity.}

For $\mathfrak{S}=+1$, \eqref{r-S-sign} is positive recalling \eqref{k-K-ineq}. Thus for this case too $r\leq 0$. However, in this case and {near the singularity $r$ can become} vanishingly small, i.e. $r\simeq 0$. This happens when $\beta$ is large ($\tanh\beta\simeq 1$) and when  $K+\kappa\simeq 1$. 

All in all, for either sign of $\mathfrak{S}$ and for any value of other parameters, $r$ is never positive. 

\item \textbf{Correlation between $\boldsymbol{f, r}$:} From \eqref{const-w-dipole-rho-X-beta} and  \eqref{F-def} one learns,  
\begin{subequations}
\begin{align}
     \frac{w}{1+w}r+f&+2 \mathfrak{S} K =-1,\label{r-f-K} \\
  {r}+f(1+w)\tanh^2\beta&=-(1+w) \left(3-2\kappa \mathfrak{S} |\tanh\beta|\right)<0 \label{r-f-ineqlty}
\end{align}
\end{subequations}
Therefore, $r,f$ can't be both positive; if $f\geq 0$ then $r<0$ and if $r\geq 0$ then  $f<0$. Recall that $r$ is never positive. In particular, if $r=0$ then $\mathfrak{S}=+1$ and $f=-1-2K$.

\item \textbf{Tilt growth and the sign of $\boldsymbol{f}$:} With the same token, one can see that $f$ does not have a definite sign; {it can have either sign depending on parameters $w, \mathfrak{S}$ and the other initial conditions}. Thus $\beta$ can grow or go to zero depending on the sign of $f$. As we will see in the examples $f$ can change sign in the course of evolution of the Universe. 
\end{enumerate}

Finally, we note that from \eqref{hubble} by inserting \eqref{const-w-dipole-rho-X-beta} and \eqref{sigma-w-dipole}, we obtain
\begin{equation}\label{H-explicit}
\begin{split}
    H^2 &=\frac{\Lambda}{3}+\frac{A_0^2}{C^2}\rho^{\frac{2w}{1+w}}\sinh^2\beta+\frac{\rho}{3}\left[1+(w+1)\sinh^2\beta\right]+ \frac{C^2 (1+w)^2}{36A_0^2} \rho^{\frac{2}{1+w}}\cosh^2\beta\\
    &= \frac{\Lambda}{3}+\frac{\rho}{3}+\frac{C^2 (1+w)^2}{36A_0^2} \rho^{\frac{2}{1+w}}+ \frac{A_0^2}{C^2}\sinh^2\beta \left[\rho^{\frac{w}{1+w}}+ \frac{C^2 (1+w)^2}{6A_0^2} \rho^{\frac{1}{1+w}}\right]^2 .
\end{split}
\end{equation}
In the first line of the above, the terms in the RHS are respectively $\Lambda$, curvature, matter and shear contributions.
We note that \eqref{h-def} is a combination of the above and \eqref{EoM-acceleration-1}.

\section{Behavior near Big Bang singularity, single fluid case}
\label{sec3:near-BB-general-w}

Analyses of the previous section are general and equations are valid for any time (any $a$). In this section, we take a closer analytical look at the near BB $a\ll 1$ regime, while the plots and numeric analysis are carried out for $a$ from BB to today $a=1$ or near future $(a\sim 100)$. {From the exponents defined in \eqref{log-log-exponents}, the Hubble parameter, density, tilt and shear can be written as} 
\begin{equation}\label{near-BB-expansion-generic-w}
\begin{split}
H=H_0 a^h,& \qquad \rho =\rho_0 a^r,\qquad \qquad H_0, \rho_0>0 \cr
\sinh\beta =\sinh\beta_0\ a^f,& \qquad \sigma =\sigma_0 a^{s-3}.
\end{split}
\end{equation}
In the $a\to 0$ limit we take $h,r,f,s$ to be ($a$ independent numbers) constants and $\beta_0, \sigma_0$ may take positive or negative values. Recalling the identity $\cosh{\beta} = \sqrt{1+\sinh^{2}{\beta}}$, the above expression for $\sinh{\beta}$  implies that near the singularity where $a \rightarrow 0$,
\begin{equation}\label{cosh-beta-T}
\cosh\beta =\left\{ \begin{array}{ccc} 1 & f>0\\ \cosh\beta_0 & f=0 \\ |\sinh\beta_0|\ a^f & f< 0 \end{array}\right. \qquad \qquad \cosh\beta\propto a^F,\quad F:= \left\{ \begin{array}{cc} 0 & \quad f \geq 0\\ f & \quad f< 0 \end{array}\right.
\end{equation}
Equations \eqref{log-log-exponents}, \eqref{r-def} and \eqref{H-explicit} in $a\ll 1$ limit yield
\begin{equation}\label{near-BB-h}
h= Min(\frac{w}{1+w}r+f, \frac{r}2+f , \frac{r}{1+w}+F)=Min (-1-2\mathfrak{S}K,\frac{r}2+f, s-3) .
\end{equation}

Next, the first line in \eqref{H-explicit} and that $\cosh^2\beta>\sinh^2\beta$ for any $\beta$ imply that, each of the curvature, matter and shear terms have a power of $\sinh\beta$ or $\cosh\beta$. As a result in either of large or vanishing $\beta$ cases the dominant term is specified by its power of $\rho$. On the other hand we note 
that $-1<w\leq 1$ and hence
\begin{equation}\label{wneq1}
\frac{2w}{1+w} \leq 1\leq \frac{2}{1+w}.    
\end{equation}
Except for the stiff matter $w=1$, where all powers of $\rho$ in $H^2$ \eqref{H-explicit} are equal to 1, shear has largest power of $\rho$, then matter and curvature the least power. 
Among  the three choices,  $r<0$, $r=0$ and $r>0$, as we argued {in the last section} $r>0$ is not possible. Since the $r=0$ and $r<0$ cases behave differently, they should be analyzed separately. {In section \ref{sec:r-ve} we explore the $r < 0$ cases for $w \neq 1$ and $w = 1$ separately and in section \ref{sec:whimper} we investigate the same for $r = 0$ cases.}
\subsection{\texorpdfstring{$r<0$}{} case}\label{sec:r-ve}
For this case  $\rho\gg 1$ and \eqref{H-explicit} implies that the $\Lambda$ term is subleading near the BB $a\ll 1$. However,  \eqref{wneq1} implies that
for $w\neq 1$ the last term in \eqref{H-explicit}, which is the shear term, dominates. 
\subsubsection{Generic \texorpdfstring{$w\neq 1$}{} case}\label{sec:r-ve-wneq1}
For this case we have
\begin{equation}\label{H-dominant-generic}
H\simeq |\sigma|/3, \qquad {\frac{\Lambda}{3}}\ll \frac{A_0^2}{a^2}e^{-4b}\ll \frac{\rho}{3}\left[1+(w+1)\sinh^2\beta\right]\ll \sigma^2/9
\end{equation}
implying
$$ \Omega_\Lambda\ll \Omega_k\ll \Omega_w\ll \Omega_\sigma\approx 1,$$
or equivalently, 
\begin{equation}\label{r<wneq1}
K=1, \qquad \kappa\ll 1.
\end{equation}
Eqs. \eqref{near-BB-h}, \eqref{sigma-w-dipole} and \eqref{r-f-K} imply 
\begin{equation}\label{hsr-fF}
\begin{split}
    h\simeq -3,\qquad s=2\kappa\tanh\beta\ll 1,\qquad r=-(1+w)(3+F),\\
 f=-1-2\mathfrak{S}+w(3+F)= \frac{1}{1-w\tanh^2\beta}(3w-1-2\mathfrak{S}),    
\end{split}
\end{equation}

\begin{figure}[!ht]
	\begin{center}
		\includegraphics[width=7.5cm]{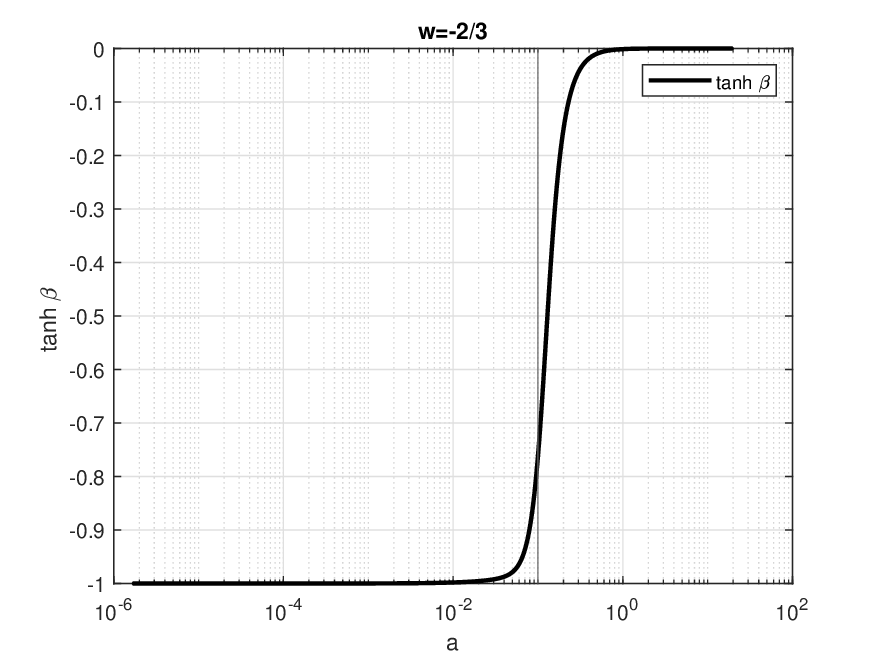}	
        \includegraphics[width=7.5cm]{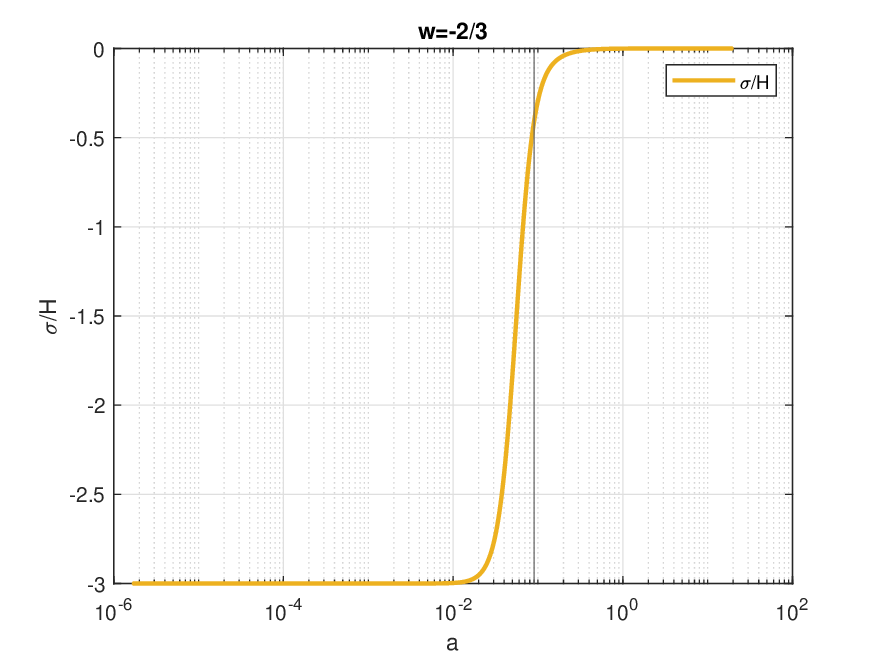}
        \includegraphics[width=7.5cm]{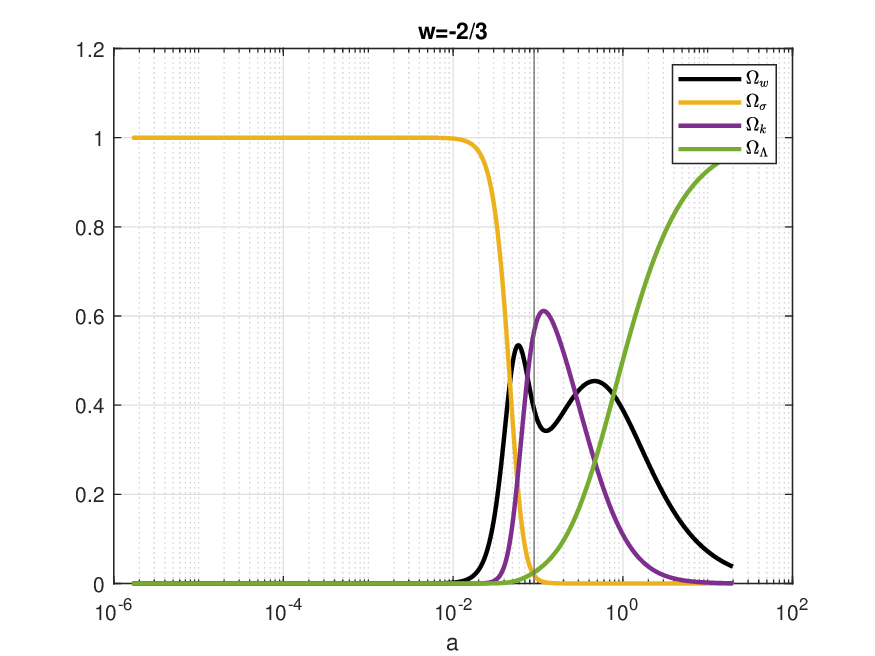}	
        \includegraphics[width=7.5cm]{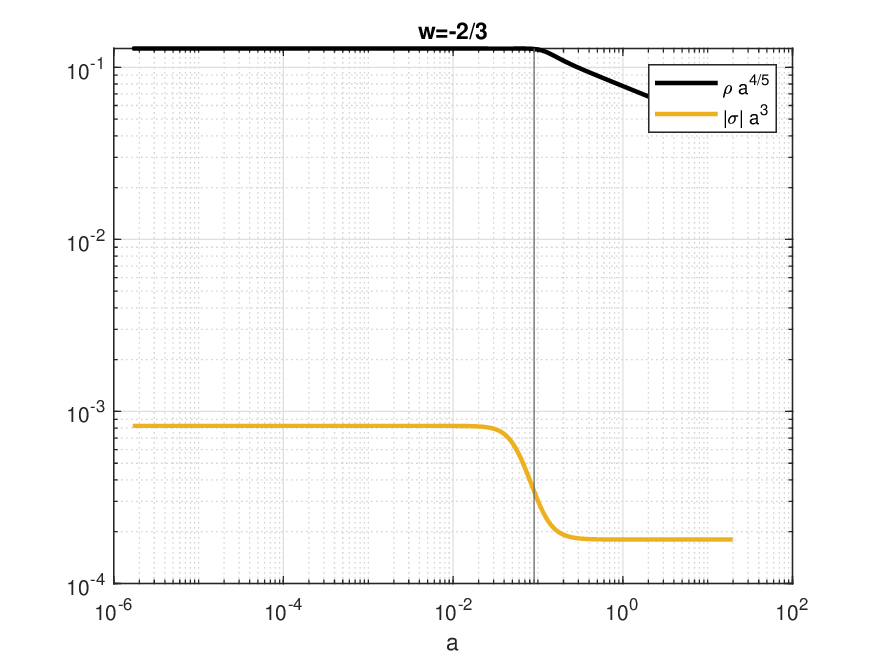}	
\caption{As an example of SEC-violating  case we show $w=-2/3$. The top-left plot shows evolution of $\beta$ which starts with a large negative value and goes to zero. The top-right plot shows $\sigma/H$ which remains $-3$ for a long time before starting to drop to zero, it  has an inflection point around $a_0\sim 0.05$. This means $K$ parameter behaves essentially like $K\simeq \tanh(a-a_0)$. The bottom-left plot shows fractional densities. It shows that the Universe starts as shear dominated, then there is a period when matter and curvature have the main contributions and are of the same order.  The details of the $\Omega_w, \Omega_k$ plots depends on the initial values. As expected in future ($a>1$) we end up with $\Lambda$ domination. Bottom-right plot shows $\rho a^{4(w+1)/(1-w)}$ and $\sigma a^3$, which are expected to be constants around $a=0$. In fact, they remain constant for a long time, till $\Omega_k, \Omega_\Lambda$ start to take over. }
		\label{Fig:w=-2/3}
	\end{center}
\end{figure}

\begin{figure}[!ht]
	\begin{center}
		\includegraphics[width=7.5cm]{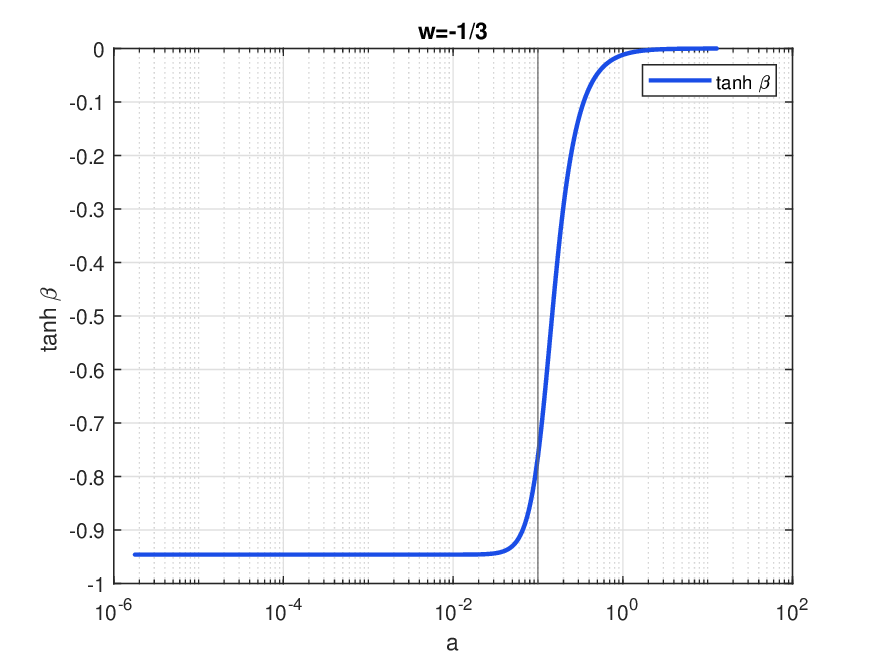}	
        \includegraphics[width=7.5cm]{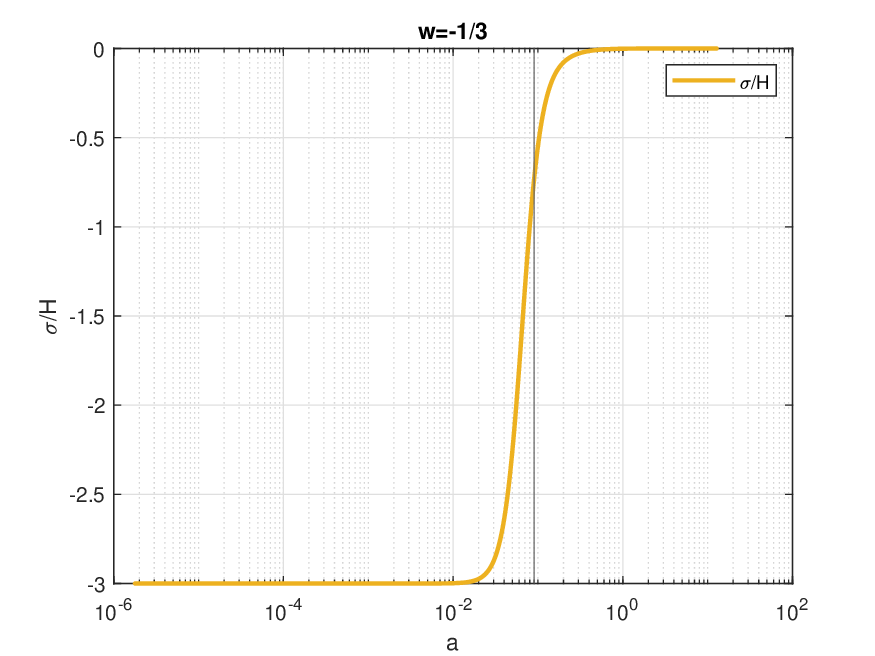}
        \includegraphics[width=7.5cm]{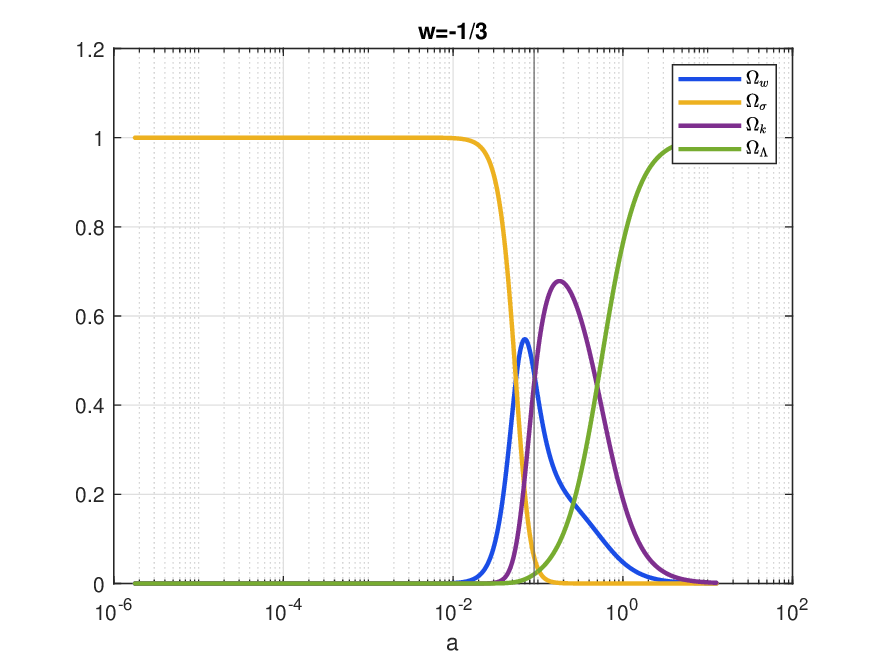}	
        \includegraphics[width=7.5cm]{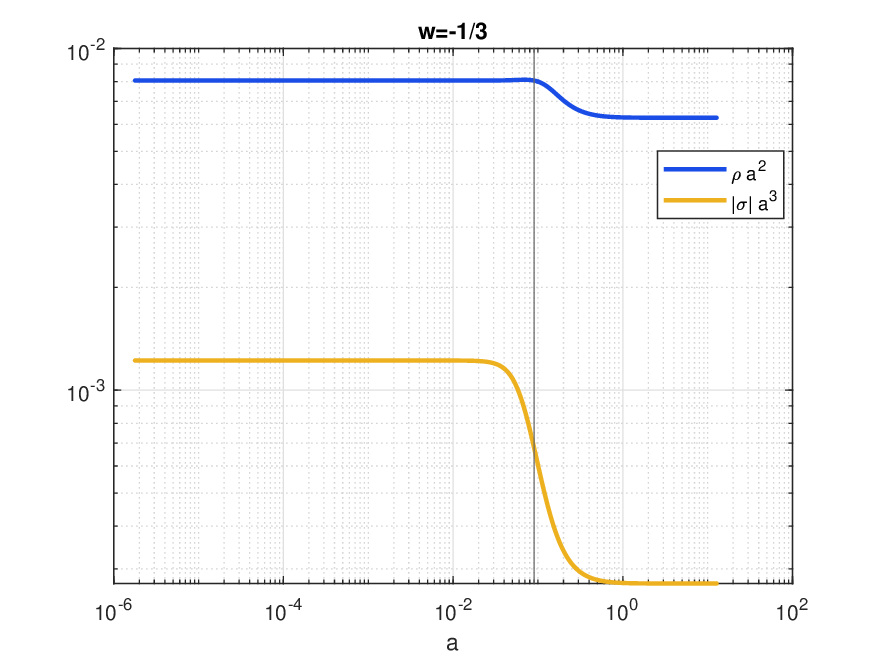}	
\caption{Plots for $w=-1/3$ case. The top-left plot shows evolution of $\tanh\beta$ which starts with a negative value and $|\beta|$ goes to zero. The top-right plot shows $\sigma/H$ which remains $-3$ for a long time before starting to drop to zero. Both $\beta$ and $\sigma/H$ plot have a $\tanh$ shape. The bottom-left plot shows fractional densities: Universe starts as shear dominated, then there is a period when matter and curvature have the main contributions and are of the same order. As expected in future ($a>1$) we end up with $\Lambda$ domination. Bottom-right plot shows $\rho a^{-3(w+1)}$ and $\sigma a^3$, which are expected to be constants around $a=0$. These two remain constant for a long time, till $\Omega_k, \Omega_\Lambda$ start to take over. When the universe has isotropized ($\sigma\sim 0$), $w=-1/3$ matter behaves like a curvature term, as in the usual FLRW setting. In this case $\rho a^2\sim const.$ at late times.}\label{Fig:w=-1/3}
	\end{center}
\end{figure}

\begin{figure}[!ht]
	\begin{center}
		\includegraphics[width=7.5cm]{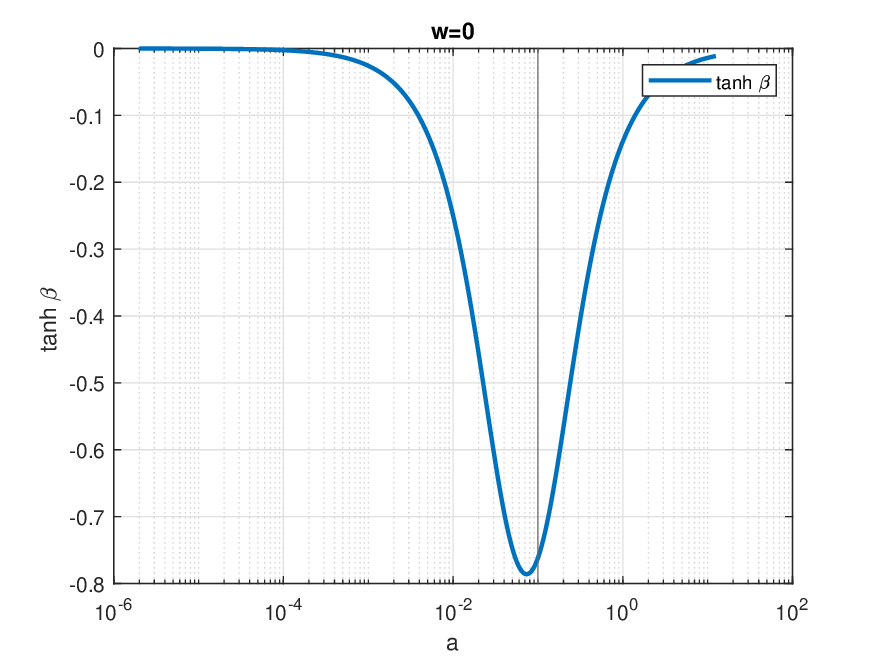}	
        \includegraphics[width=7.5cm]{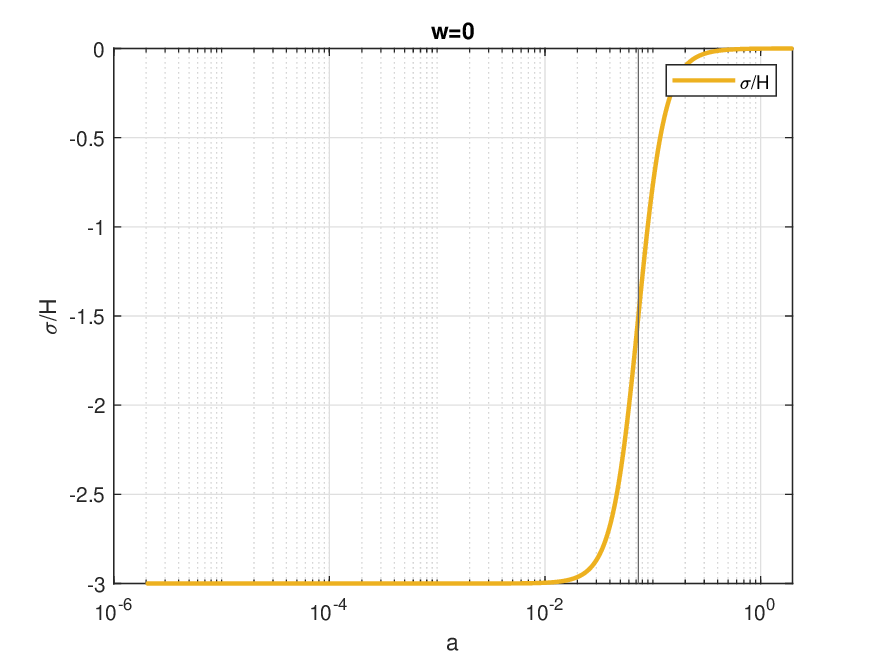}
        \includegraphics[width=7.5cm]{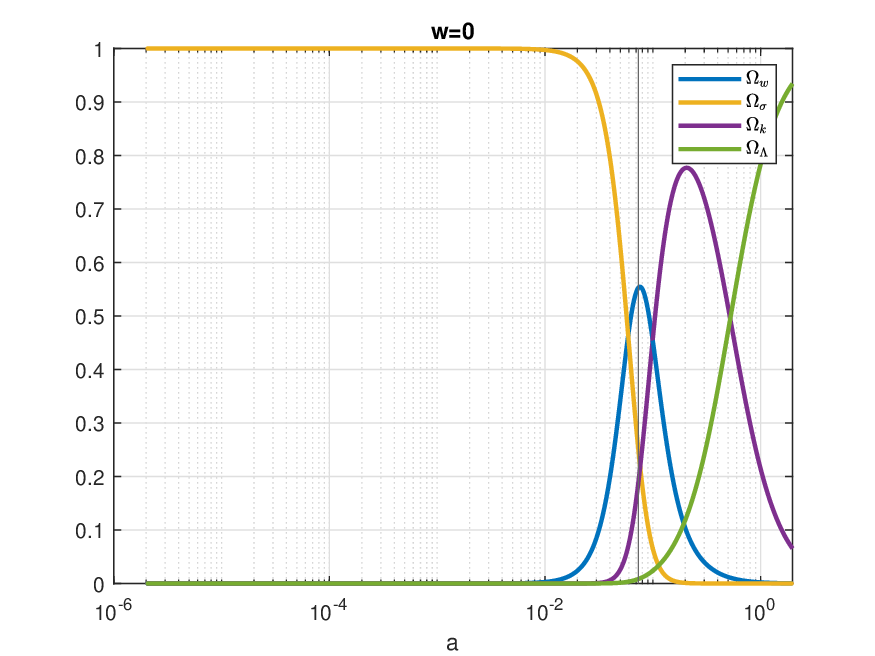}	
        \includegraphics[width=7.5cm]{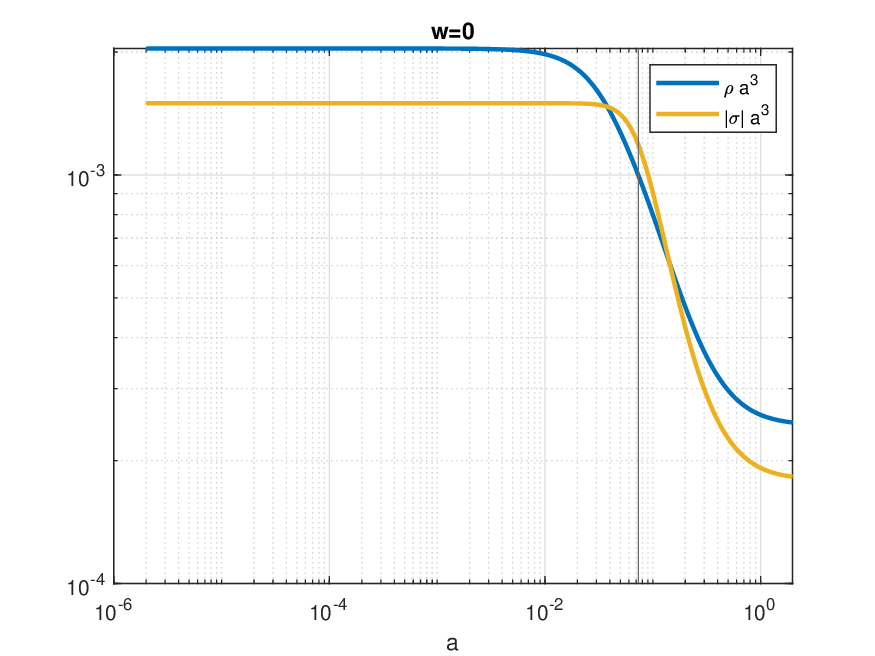}	
\caption{As an example of $w>-1/3$ here we show  $\omega=0$ case. The top-left plot shows evolution of $\beta$ which starts with a negative value and $|\beta|$ grows until it reaches a maximum value. The top-right plot shows $\sigma/H$ which remains $-3$ for a long time before a quick drop to zero.  The bottom-left plot shows  that the Universe starts as shear dominated, then there is a very brief matter dominated era, before settling to a curvature dominated regime. As expected in future ($a>1$) we end up with $\Lambda$ domination. The inflection point of $\sigma/H$, the minimum of $\beta$ and the maximum point for $\Omega_w$  happen at the same time. The Bottom-right plot shows $\rho a^{3(w+1)}$ and $\sigma a^3$, which are expected to be constants around $a=0$. In fact, they remain constant for a long time, till $\Omega_k, \Omega_\Lambda$ start to take over.}\label{Fig:w=0}
	\end{center}
\end{figure}

\begin{figure}[!ht]
	\begin{center}
		\includegraphics[width=7.5cm]{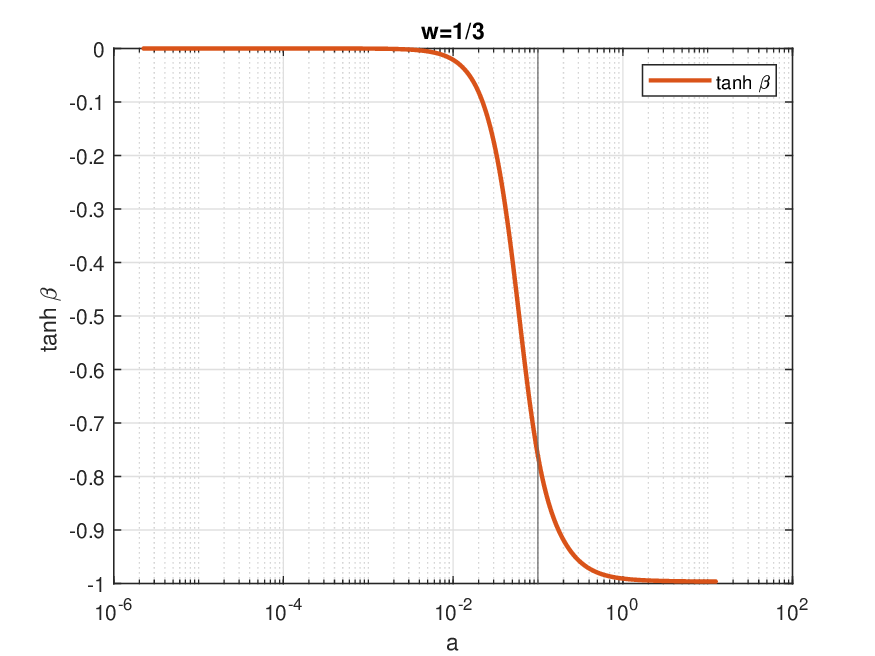}	
        \includegraphics[width=7.5cm]{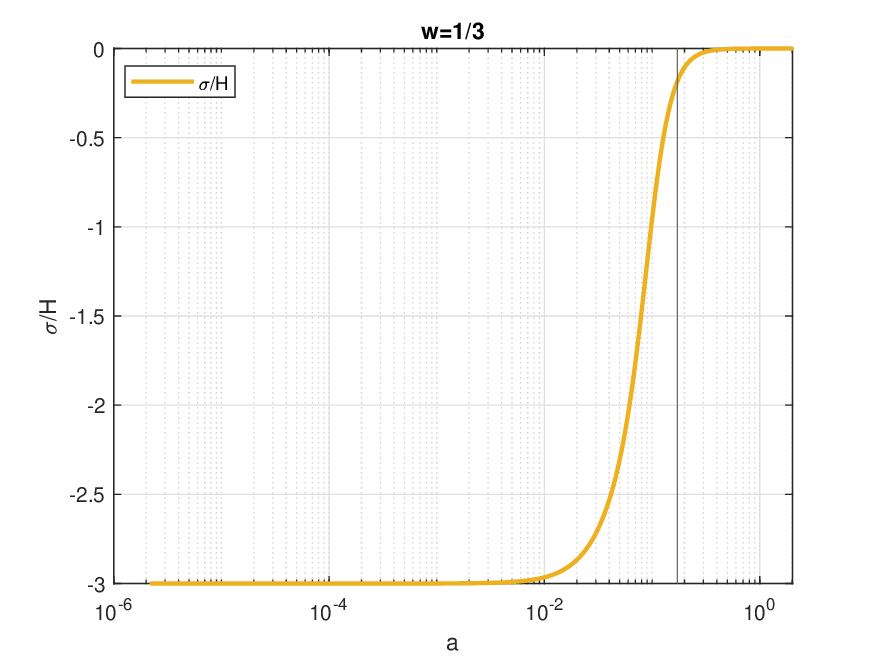}
        \includegraphics[width=7.5cm]{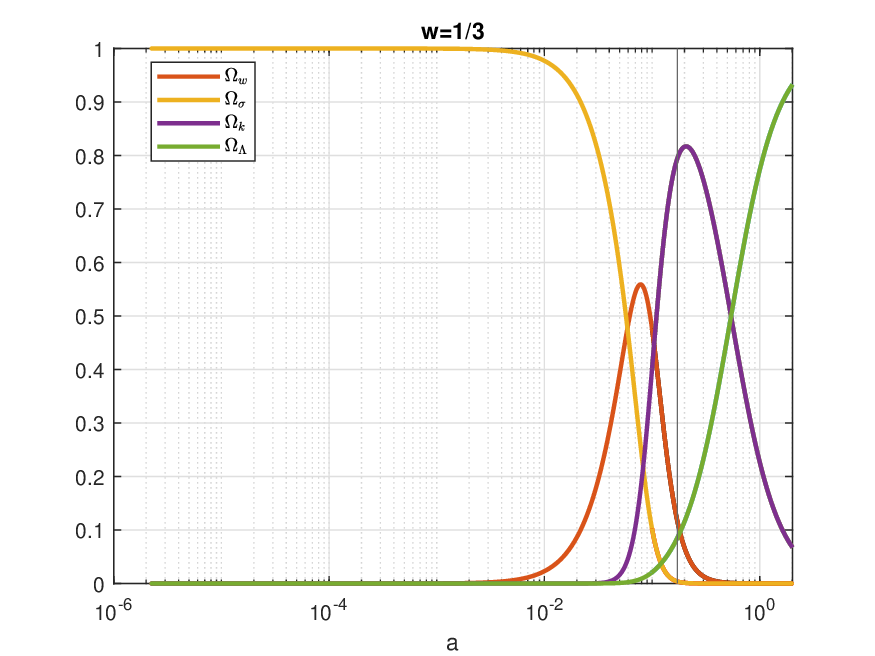}	
        \includegraphics[width=7.5cm]{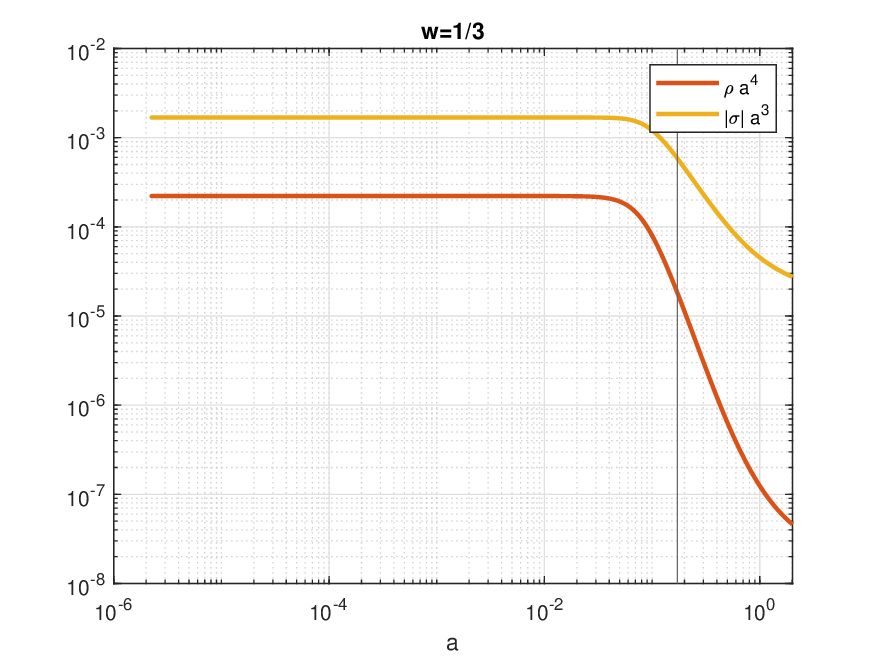}	
\caption{As an example of SEC-respecting case we present $w=1/3$. The top-left plot shows evolution of $\beta$ which starts with a negative value and $|\beta|$ grows to very large values. This is in accord with the analysis in \cite{Ebrahimian:2023svi}. The top-right plot shows $\sigma/H$ which remains $-3$ for a long time before a rapid drop to zero. The bottom-left plot shows  that the Universe starts as shear dominated, then there is a matter dominated era, before settling to a curvature dominated regime. As expected in future ($a>1$) we end up with $\Lambda$ domination. Bottom-right plot shows $\rho a^{3(w+1)}$ and $\sigma a^3$, which are expected to be constants around $a=0$. }
		\label{Fig:w=+1/3}
	\end{center}
\end{figure}

From the second line in \eqref{hsr-fF}, and recalling the definition of $F$ in \eqref{cosh-beta-T}, one learns that for $r<0$, $\mathfrak{S}=-1$. These also imply that the Strong Energy Condition (SEC) respecting and SEC-violating, respectively  $w>-1/3$ and $-1<w\leq -1/3$ cases, behave differently. So, we discuss these cases separately. 
First we note that  
\begin{equation}\label{r-f-r-ve}
\begin{split}
   f=3w+1,\quad r=-3(1+w) \qquad &\text{for} \qquad   -1/3\leq w <1 \\
    f=\frac{3w+1}{1-w},\quad r=-4\left(\frac{1+w}{1-w}\right)     \qquad &\text{for}  \qquad -1<w\leq -1/3
\end{split}
\end{equation}
For $r<0$  matter density $\rho$ grows near the BB and \eqref{r-f-r-ve} implies that 
\begin{itemize}
    \item for SEC-violating case $w<-1/3$, $f<0$ and tilt can grow;\item for $w=-1/3$ tilt remains constant and
    \item for SEC-respecting case of $w>-1/3$ tilt dies off.
\end{itemize}
In Figs.~\ref{Fig:w=-2/3}, \ref{Fig:w=-1/3}, \ref{Fig:w=0} and \ref{Fig:w=+1/3} we have depicted evolution of a Universe with respectively $w=-2/3, w=-1/3, w=0, w=1/3$ with some generic initial conditions. Note that while the discussions in this section, unless mentioned explicitly, are for near Big Bang $a\ll 1$, the figures show the course of the evolution of Universe from very small $a$ to near future $a\sim 100$.  For $0 < w < 1$ cases,  we are consistent with the results in \cite{COLLINS:1979} for the near BB behavior, however, our numeric analyses allow us to also study and plot the behavior of the system away from the BB. 

Next, we consider \eqref{const-w-dipole-rho-X-beta} and \eqref{sigma-w-dipole}, yielding
\begin{equation}\label{rho0beta0}
\begin{split}
 C&=\rho_0^{\frac{w}{1+w}}\sinh\beta_0,\\
\frac{6}{1+w} A_0 H_0&= \left\{\begin{array}{ccc} -\rho_0\sinh\beta_0  \quad \text{for}\quad  -1/3 <w<1 \\ -\rho_0\sinh\beta_0 \cosh\beta_0 \quad \text{for}\quad w= -1/3 \\ -\rho_0\sinh\beta_0|\sinh\beta_0| \quad \text{for} \quad -1<w< -1/3 \end{array}\right.
\end{split}
\end{equation}
That is $\beta_0<0$ and also shear is negative. 

\paragraph{Summary of {near  Big Bang ($a\ll 1$) behavior} for $r<0$ and $-1<w<1$ case:} 
\begin{equation}
    H=H_0a^{-3},\quad \sigma=-3H_0 a^{-3},\quad a e^b=1, \quad 
    \rho\sinh2\beta=-\frac{12H_0 A_0}{1+w} \ a^{-2}
\end{equation}
and 
\begin{equation}
\begin{split}
   \rho=\rho_0 a^{-3(1+w)}, \qquad \sinh\beta=\sinh\beta_0 a^{3w+1}, \qquad &\text{for} \qquad   -1/3\leq w <1 \\
   \rho=\rho_0 a^{-4(1+w)/(1-w)}, \qquad \sinh\beta=\sinh\beta_0 a^{\frac{3w+1}{1-w}},      \qquad &\text{for}  \qquad -1<w\leq -1/3
\end{split}
\end{equation}
where $\rho_0, \beta_0, H_0$ are related as in \eqref{rho0beta0}. 

Some remarks are in order.
\begin{enumerate}
\item The universe is decelerating in this case as $\ddot{a}/a=-2H^2<0$.
\item Big Bang is shear dominated  and the sign of $\sigma$ and $\beta$ are both negative. 
\item Tilt is not necessarily vanishing near BB: while for SEC-respecting $w>-1/3$ case it vanishes, for $w=-1/3$ it remains constant and for SEC-violating $w< -1/3$ case $|\beta|$ grows around $a=0$. This result matches those in \cite{Hewitt:1992sk}.
\item We note that
\begin{equation}
\begin{split}
  \Omega_\Lambda\sim a^{6},\quad \Omega_k\sim a^{8},\quad \Omega_w\sim a^{4},\quad \Omega_\sigma\simeq 1,  \quad &\text{for}  \quad -1<w\leq -1/3\\  
  \Omega_\Lambda\sim a^{6},\quad \Omega_k\sim a^{8},\quad \Omega_w\sim a^{3(1-w)},\quad \Omega_\sigma\simeq 1, \quad &\text{for} \quad   -1/3\leq w <1.
   \end{split}
\end{equation}
\item In the usual FLRW case, spatial curvature effectively behaves as a matter with $w=-1/3$. However, as the above equations show, this is not the case for the dipole cosmology; for $w=-1/3$, $\Omega_w\sim a^{4}$ whereas $\Omega_k\sim a^8$. This difference could be traced to the anisotropy, recalling \eqref{Omega-def} and that $\Omega_k$ has a factor of $e^{4b}$. Note also that for $w=-1/3$, $\sinh^2\beta$ is essentially a constant. 
\item In terms of cosmic comoving time,
\begin{equation}
    a^3=3H_0 t,\quad H=\frac{1}{3t},\quad \rho\sinh2\beta=-\frac{6A_0 H_0 }{1+w} (3H_0 t)^{-2/3}
\end{equation}
\item $a(t), b(t)$, and the $t$ power of $\rho\sinh2\beta$ do not depend on the EoS $w$. 
\item  $T^0_z=-6A_0 H$  is independent of $w$ and other initial values and is pure geometric.
\item Metric near the BB takes the form
\begin{equation}\label{nearBB-metric-generic}
\begin{split}
    ds^2&=-dt^2+\frac{1}{a^2}dz^2+a^4 e^{-2 A_0 z}(dx^2+dy^2)\\
    &=-dt^2+t^{-\frac{2}{3}}dz^2+t^{\frac{4}{3}} e^{-2 A_0 z}(dx^2+dy^2)
\end{split}
\end{equation}
That the above metric is independent of cosmological constant $\Lambda$, $w$ and $\beta_0, \rho_0$ is a manifestation of John Wheeler's ``near the singularity, matter does not matter.'' 
\item The proper distances along $z$ direction are respectively increasing and decreasing along $x$ and $y$ directions as we reach $a\to 0$.  We dub this as \textit{filament Universe} in which the universe starts effectively in a phase with $1+1$ dimensions and expands to $1+3$ dimensions. 
\item We comment in passing that \eqref{nearBB-metric-generic} is an ``open Kasner'' geometry in which constant time slices is an open Universe, a hyperboloid of radius $1/A_0$.
\item After the initial anisotropic phase, the universe evolves to an isotropic phase in almost one Hubble time, $t=1/(3H_0)$, when $a\approx 1$, $b\approx 0$.

\end{enumerate}

\paragraph{Singularity structure.}  From Einstein equations we learn that
\begin{equation}\label{R-T-R2-T2}
    R=-T=\rho-3p,\qquad R_{\mu\nu}R^{\mu\nu}=T_{\mu\nu}T^{\mu\nu}=\rho^2+3p^2, 
\end{equation}
where we used \eqref{T-trace-T2}. Therefore, Ricci scalar and Ricci tensor-squared are respectively diverging as $a^{r}$ and $a^{2r}$. Moreover, one can see that any scalar made of Ricci curvature is $\beta$ independent and if its maximum power of $R$ is $p$, it diverges as $a^{rp}$.\footnote{Note that \eqref{nearBB-metric-generic} gives the leading behaviour of the metric near $a=0$. To obtain \eqref{R-T-R2-T2}, which is based on Einstein equations one should also consider subdominant terms in the metric.} Besides the Ricci and scalars built from it, one may explore Weyl tensor and its divergence. 
The electric  Weyl tensor for this metric has the components (cf. \eqref{Electric-Weyl})\footnote{If we use the fluid comoving frame with the fluid four-velocity \eqref{u-mu}, we find the the electric Weyl has the components 
\begin{align}
 \Tilde{\mathcal{E}}^x_x= \Tilde{\mathcal{E}}^y_y= {\mathcal{E}}^x_x,\qquad 
\Tilde{\mathcal{E}}^t_t=   {\mathcal{E}}^x_x \sinh^2\beta, \qquad 
\Tilde{\mathcal{E}}^z_z=-{\mathcal{E}}^x_x\cosh^2\beta,\cr
\Tilde{\mathcal{E}}^t_z=-a^{2 K\mathfrak{S}+1}\cosh\beta\sinh\beta {\mathcal{E}}^x_x,\qquad 
\Tilde{\mathcal{E}}^z_t=a^{-2 K\mathfrak{S}-1}\cosh\beta\sinh\beta {\mathcal{E}}^x_x\nonumber
\end{align}
where ${\mathcal{E}}^x_x$ is given in \eqref{Electric-Weyl-2}.}
\begin{equation}\label{Electric-Weyl-2}
    \mathcal{E}^x_{\;x}=\mathcal{E}^y_{\;y}=-\frac12\mathcal{E}^z_{\;z}
   \simeq 2K(-\mathfrak{S}+K)H^2\sim a^{-6}.
\end{equation}
and as a consequence of homogeneity of our metric, the magnetic part of Weyl vanishes. Recalling \eqref{r-f-r-ve} and that $r>-3$, we observe that  Ricci tensor invariants are subdominant compared to Weyl invariants of the same curvature power, i.e. we have Weyl dominance in the terminology of \cite{Barrow:2002is}. This is in contrast to the usual FLRW cosmology where Ricci invariants dominate (and Weyl curvature is zero). This discussion of singularity behavior of curvature invariants confirms the picture of filament Universe mentioned above.

 { Observe that \eqref{nearBB-metric-generic} is time symmetric. Therefore, if there exists a big crunch singularity, our analysis above applies there too. }

\subsubsection{Stiff matter \texorpdfstring{$w=1$}{} case}\label{sec:r-ve-stiff-matter}

In this case, 
\begin{equation}\label{H-explicit-w=1}
    H^2\simeq \frac\rho3(1+\frac{C^2}{3A_0^2})+\rho\sinh^2\beta(\frac{A_0}{C}+\frac{C}{3A_0})^2 .
\end{equation}
It appears that $f>0$, $f=0$ and $f<0$ should be analyzed separately for $w=1$. Straightforward analysis shows that $f\leq 0$ cases are not possible and one remains with $f>0$ case, for which\footnote{Equations \eqref{r-def} and \eqref{f-def} imply that $\rho \sinh^2\beta=C^2 (a e^{2b})^{-2}$ not only for near BB, but during the whole evolution of the Universe.}
\begin{equation}
    H=H_0a^{-3},\quad \sigma=3\mathfrak{S} K H,\quad e^b=a^{\mathfrak{S}K},\quad \rho=\rho_0 a^{-6},\quad  \sinh\beta=\sinh\beta_0 a^{2-2\mathfrak{S}K}, 
\end{equation}
where 
\begin{equation}
    K=(1+\frac{3A_0^2}{C^2})^{-1/2},\quad \rho_0\sinh^2\beta_0=C^2,\quad \rho_0={3H_0^2}(1-K^2).
\end{equation}
Therefore, for either sign of $\mathfrak{S}$, tilt vanishes near the BB while the density diverges as $a^{-6}$.

\begin{figure}[!ht]
	\begin{center}
		\includegraphics[width=7.5cm]{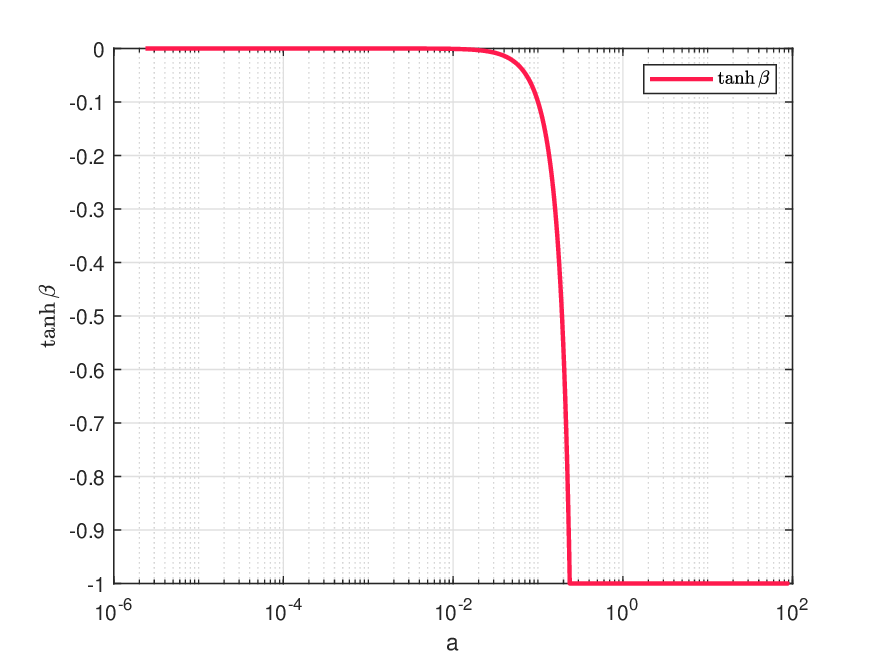}	
        \includegraphics[width=7.5cm]{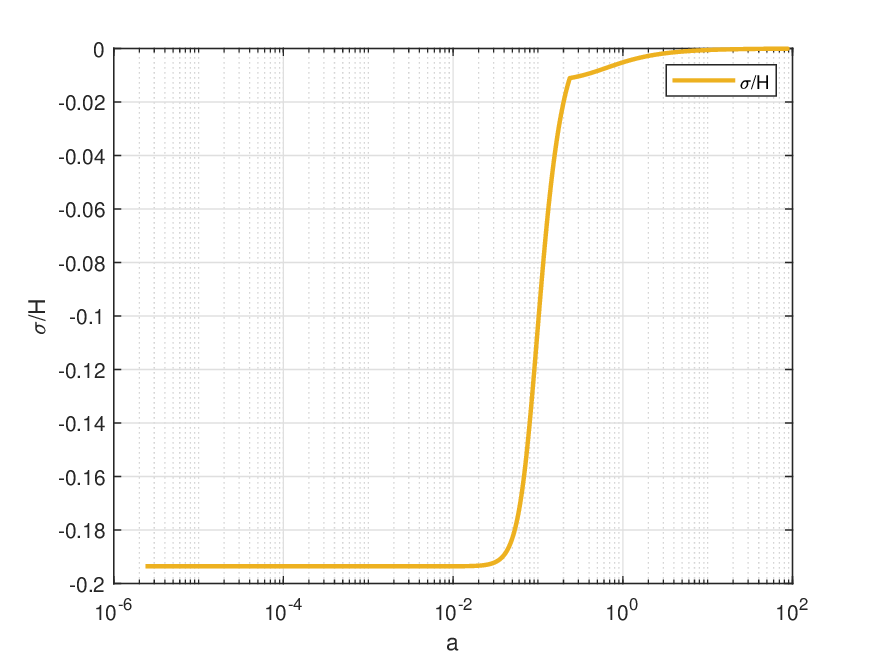}
        \includegraphics[width=7.5cm]{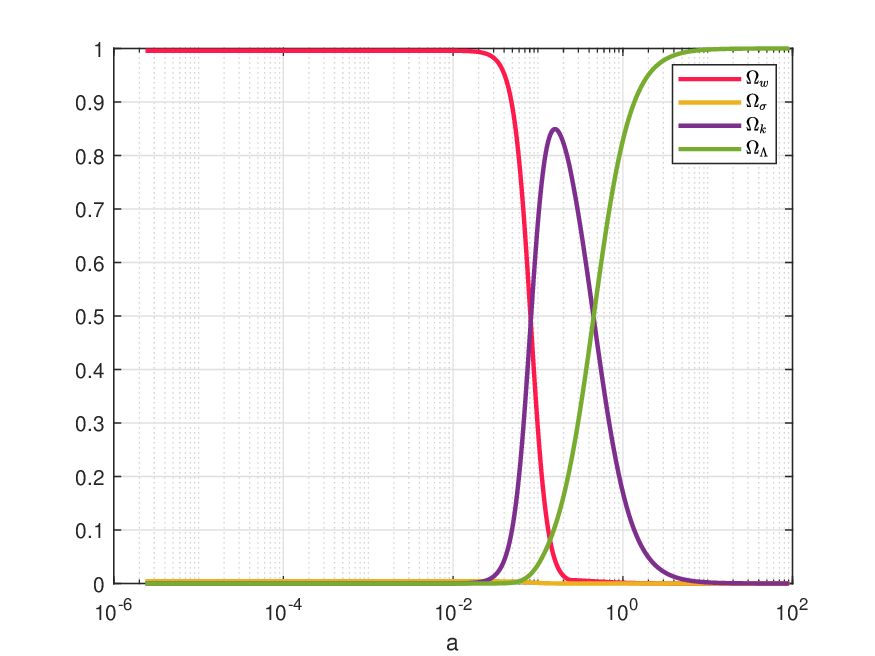}	
        \includegraphics[width=7.5cm]{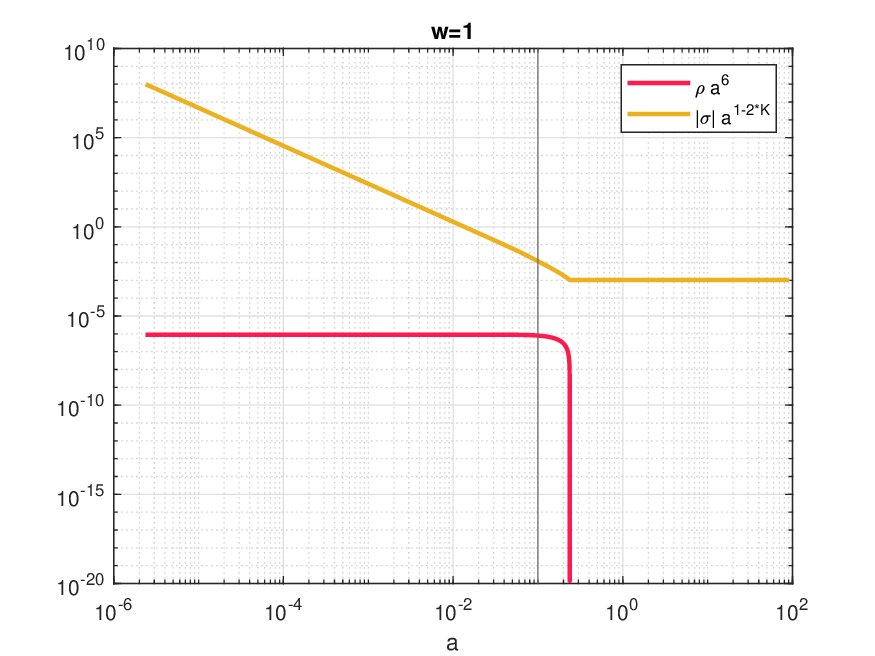}	
\caption{Plots for stiff matter $w=1$, $\beta<0$ ($\mathfrak{S}=-1$) case. As we see $|\tanh\beta|$ reaches its maximal value $1$ and stays there and the Universe isotropises fast while $\Omega_k$ can become sizable in the meantime.  }
		\label{Fig:stiffnegative}
	\end{center}
\end{figure}

\begin{figure}[!ht]
	\begin{center}
		\includegraphics[width=7.5cm]{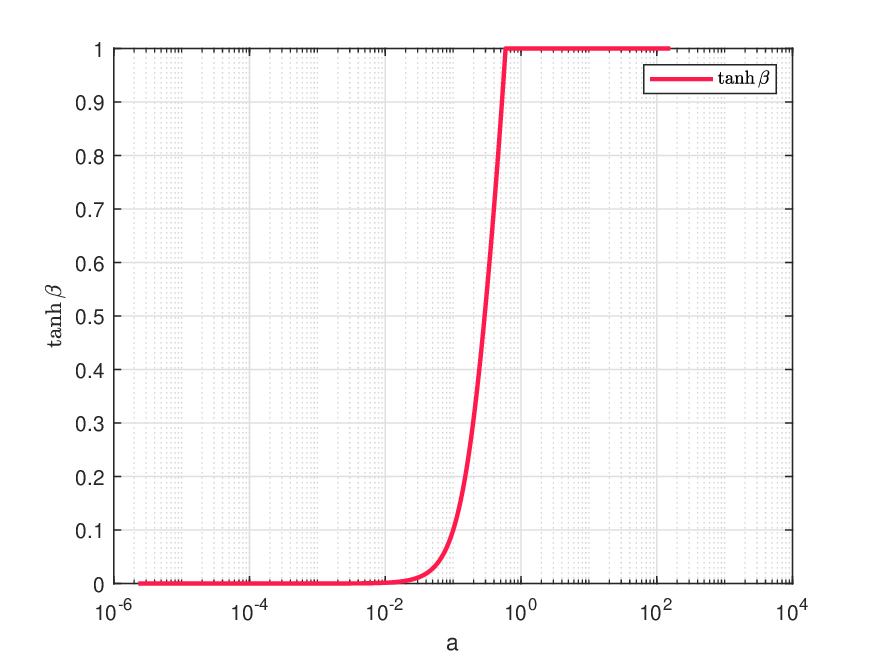}	
        \includegraphics[width=7.5cm]{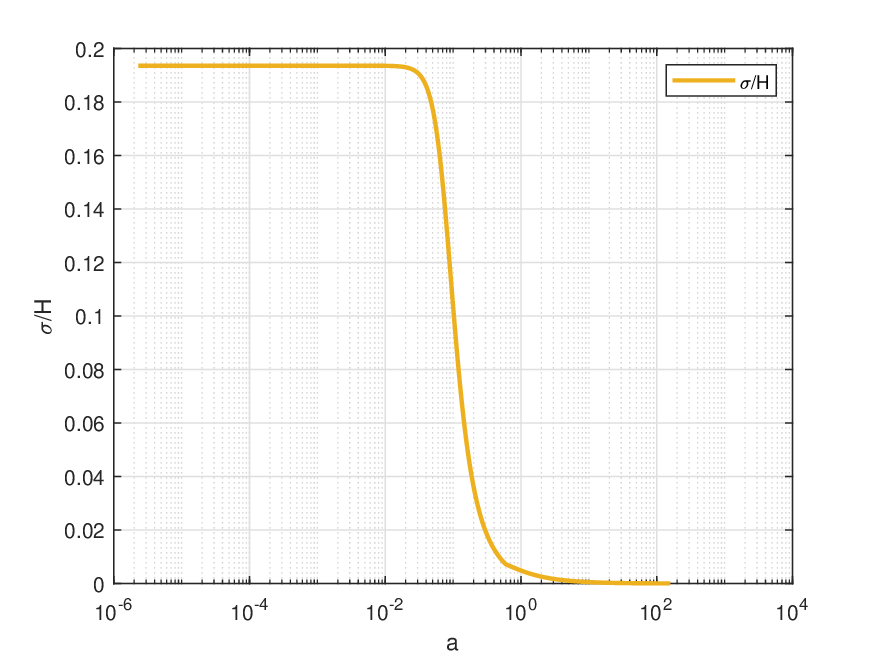}
        \includegraphics[width=7.5cm]{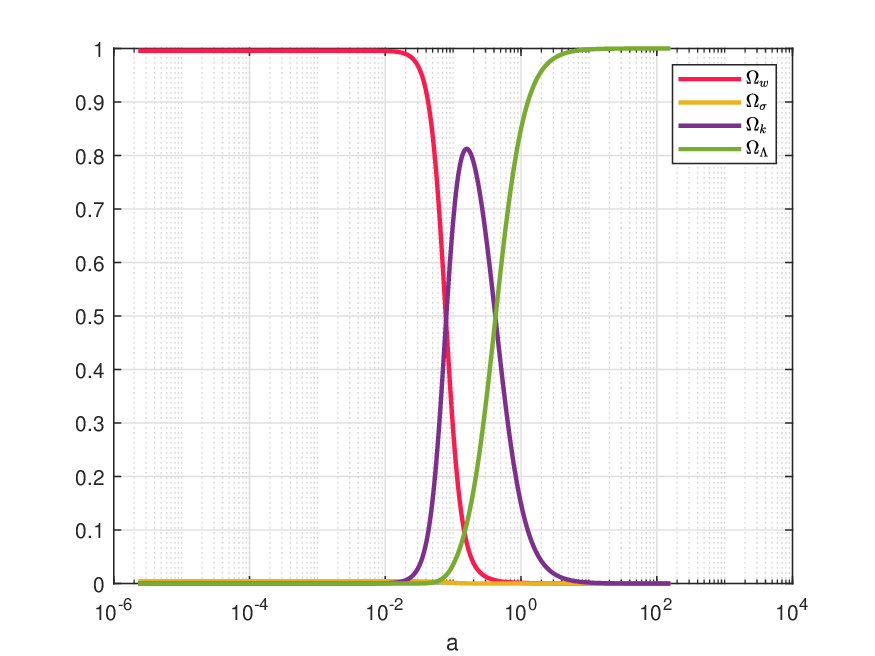}	
        \includegraphics[width=7.5cm]{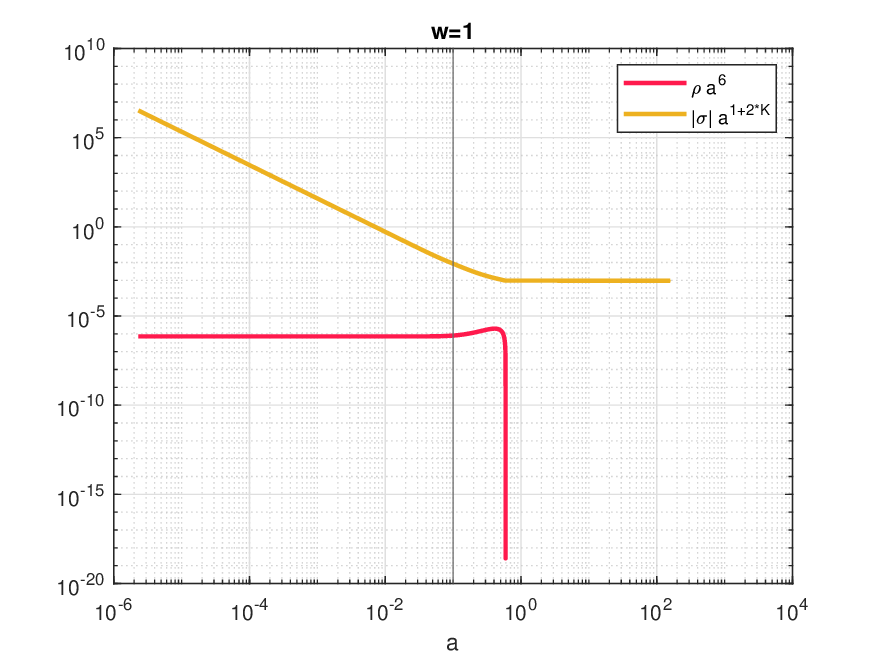}	
\caption{Plots for stiff matter $w=1$,  $\beta>0$ ($\mathfrak{S}=+1$) $\Lambda\neq 0$ case. As we see $\tanh\beta$ increases to its maximal value $1$ very fast and stays there while the Universe isotropizes fast. $\Omega_k$ can become sizable in the meantime, before being taken over by $\Omega_\Lambda$ at late times.}
		\label{Fig:stiffpositive-Lambda}
	\end{center}
\end{figure}

\begin{figure}[!ht]
	\begin{center}
		\includegraphics[width=7.5cm]{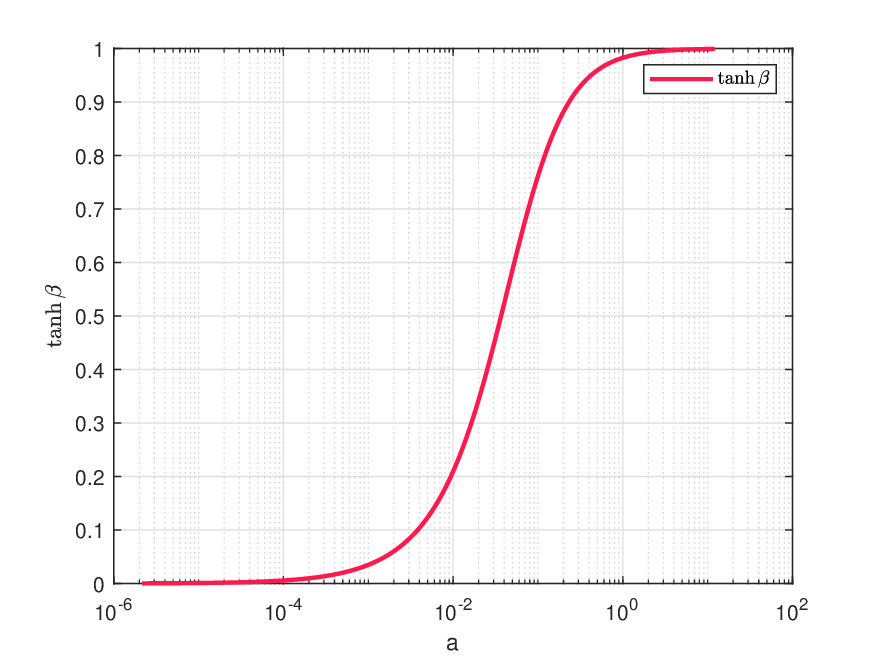}	
        \includegraphics[width=7.5cm]{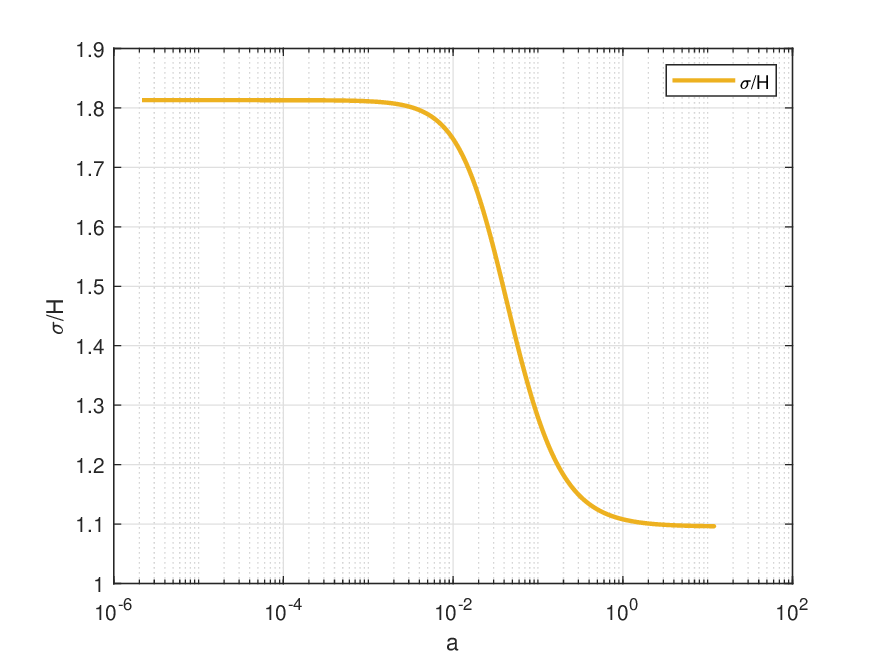}
        \includegraphics[width=7.5cm]{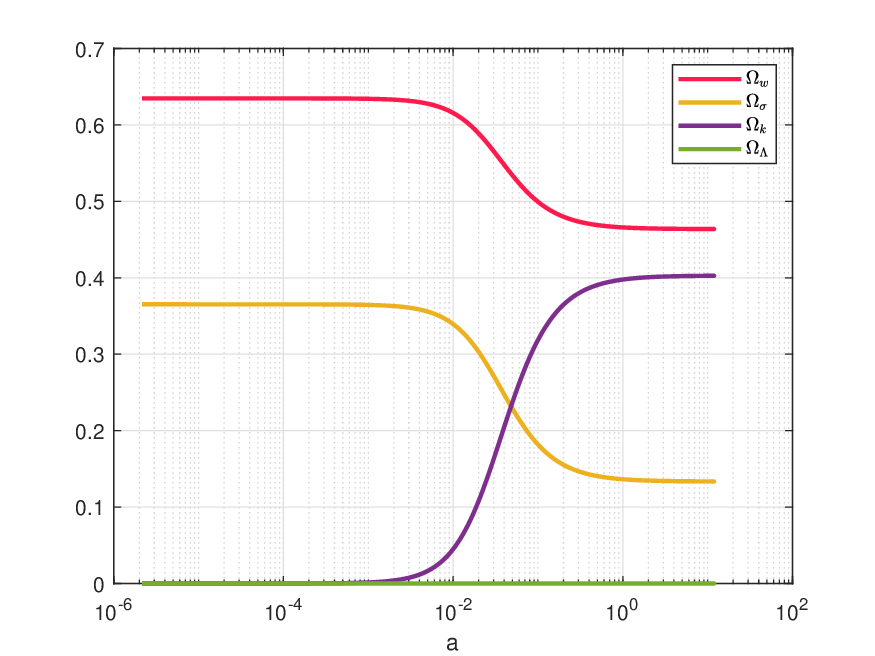}	
        \includegraphics[width=7.5cm]{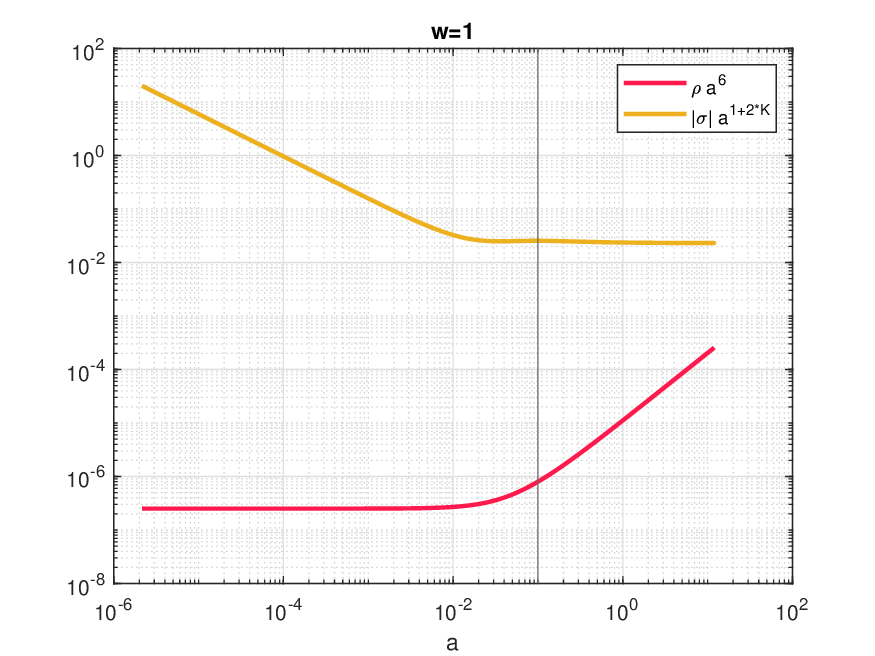}	
\caption{Plots for stiff matter $w=1$,  $\beta>0$ ($\mathfrak{S}=+1$) case $\Lambda=0$. While $\tanh\beta$ grows to its maximal value $1$, it does so milder than the $\Lambda\neq 0$ case shown in Fig.~\ref{Fig:stiffpositive-Lambda}. Moreover, in this case in asymptotic future $\sigma/H$  and $\Omega_k, \Omega_w, \Omega_\Lambda$ remain finite, i.e. the Universe does not isotropize. }
		\label{Fig:stiffpositive-no-Lambda}
	\end{center}
\end{figure}

Since $K^2<1$, we do not consider $K=1$ case which corresponds to essentially empty space ($\rho_0\sim 0$). 
For this case the shear and the matter contributions are of the same order and 
\begin{equation}
{\Omega_\sigma}=\frac{\sigma^2/9}{H^2}= K^2=1-{\Omega_{w=1}}, \qquad \Omega_\kappa\sim a^{4(1-\mathfrak{S}K)}, \qquad \Omega_\Lambda\sim a^6. 
\end{equation}
In this case sign of $\sigma$ (and/or $\beta$) $\mathfrak{S}$ is not fixed. Evolution of the Universe for $\mathfrak{S}=\pm 1$ cases has been depicted in Figs.~\ref{Fig:stiffnegative}, \ref{Fig:stiffpositive-Lambda} and \ref{Fig:stiffpositive-no-Lambda}.

For this case, metric at early times takes the form
\begin{equation}\label{nearBB-metric-stiff-matter}
\begin{split}
    ds^2&=-dt^2+a^{4\mathfrak{S}K+2}dz^2+a^{2-2\mathfrak{S}K} e^{-2 A_0 z}(dx^2+dy^2)\\
    &=-dt^2+t^{\frac{4\mathfrak{S}K+2}{3}}dz^2+t^{\frac{2-2\mathfrak{S}K}{3}} e^{-2 A_0 z}(dx^2+dy^2).
\end{split}
\end{equation}
Unlike the generic $w$ case, this is not an open Kasner type solution, as it is not a vacuum solution and the matter effects appearing through $K$, are of the same order as the shear. 
For $\mathfrak{S}=+1$, the combinations $4K+2, 2-2K$ are both positive and physical length in all spatial directions shrinks to zero, but anisotropically. For $\mathfrak{S}=-1$, $2+2K >0$ while $2-4K$, depending on the value of $K$ can have positive or negative signs. For $K<1/2$ scale factors along all spatial directions go to zero, whereas for $K>1/2$ we have a filament like Universe near the BB. See \cite{Erickson:2003zm} for more discussions. {Finally, we comment that for $K=1/2$ we recover the ``Barrel Singularity'' behavior as discussed in \cite{COLLINS:1979}.}

\paragraph{Singularity structure.} In this case matter and Ricci scalar both blow up like $a^{-6}$ and hence we have a usual Big Bang singularity (in the sense that Ricci scalar blows up). All invariants built from Ricci curvature also blow up, a  power $p$ Ricci-curvature invariant blows up  at $a^{-6p}$.

The electric Weyl tensor has the same components as in \eqref{Electric-Weyl-2}, while magnetic components of the Weyl tensor vanish. We see that Weyl curvature and Ricci curvature invariants behave in the same way, a power $p$ invariant blows up like $a^{-6p}$. In the terminology of \cite{Barrow:2002is}, this is an example of Weyl-Ricci balanced case.

The nonzero components of Riemann (cf. \eqref{Riemann-frame}) take the form
\begin{subequations}\begin{align}
    \Tilde{R}_{txtx}= \Tilde{R}_{tyty}\approx-H^2(K-\mathfrak{S})(K+2\mathfrak{S})\\
    \Tilde{R}_{tztz}\approx -2H^2(K-\mathfrak{S})(2 K+\mathfrak{S})    
\end{align}\end{subequations}
If we  exclud empty space case of $K=1$, $\Tilde{R}_{txtx}$ is positive (negative) for negative (positive)  $\mathfrak{S}$. While $\Tilde{R}_{tztz}$ is always negative for $\mathfrak{S}=+1$, for $\mathfrak{S}=-1$ it is positive (negative) for $K<1/2$ ($K>1/2$) and vanishes for $K=1/2$. This fact dovetails with the above comment about the metric (cf. discussions below \eqref{nearBB-metric-stiff-matter}).

\paragraph{$w=1$ evolution, away from BB.} As discussed in \cite{Krishnan:2022qbv, Krishnan:2022uar} for $\mathfrak{S}=+1$ and $w>1/3$ tilt grows in future. In principle this growth can continue indefinitely in future along with the expansion of the Universe. Moreover, presence of $\Lambda$, which dominates the late and future of the Universe, will impact the tilt growth and may even yield a new kind of behavior at a finite time (finite $a$). In this part we explore the $w=1$ at larger $a$ and with or without $\Lambda$ more closely. In this case we have
\begin{equation}
\begin{split}
    f&=2\cosh^2\beta \left[1-\mathfrak{S}(K+\kappa |\tanh\beta|)\right], \\ 
    r &=-2-4\cosh^2\beta(1-\mathfrak{S}|\tanh\beta|(\kappa+K|\tanh\beta|).
\end{split}
\end{equation}
From \eqref{k-K-ineq} we learn that $K+\kappa |\tanh\beta|<1$ and hence $f>0$ for either sign of $\mathfrak{S}$. Similarly one also learns that $r<-2$ for either sign of $\mathfrak{S}$, i.e. $\beta$ grows (indefinitely) in asymptotic future and the matter to dilute away ($\rho\to 0$) in future. This is compatible with the results in \cite{Krishnan:2022uar} where it is shown that for $w>1/3$ the tilt grows in far future. On the other hand recalling that $r/2+f=-1-\mathfrak{S}K$ can have either signs depending $\mathfrak{S}$.

\paragraph{$\bullet\ {\mathbf{\mathfrak{S}=-1}}$.} In this case $f>0$ and $r<0$ for the whole course of evolution and hence $f$ is a monotonically  increasing function of $|\beta|$ and  $r$ is a monotonically decreasing function of $|\beta|$. This means $f\geq 2+2K$ and $r\leq -6$. So, $\beta$ grows at large $a$ and  matter energy density dilutes away.  
For this case we expect a very gradual $\beta$ growth at early times, nonetheless since $f$ can basically become very large for large $\beta$, $|\beta|$ can in principle become very large very fast at a finite $a$. For large enough $\beta$, $f\simeq 2\cosh^2\beta$ and one can integrate \eqref{f-def} to obtain $\tanh\beta/\tanh\beta_0= (a/a_0)^2$, which means $\beta\to\infty$ for a finite $a$ and $\rho$ is just simply decreasing. Therefore, the ``singularity'' in $\beta$ at finite $a$ is not a curvature singularity. We note that for large $\beta$,  $\Omega_w=2\kappa K, \Omega_\sigma=K^2, \Omega_k=\kappa^2$ can all be of order one at large $a$. However, in far enough future eventually $\Lambda$ takes over, $K, \kappa$ tend to zero and we end up with an isotropic almost de Sitter Universe. The evolution for this case is depicted in Fig.~\ref{Fig:stiffnegative}.

\paragraph{$\bullet\  {\mathbf{\mathfrak{S}=+1}}$.} In this case $f>0$ and $r<-2$ and hence $r/2+f=-2K-1<0$. Compared to the $\mathfrak{S}=-1$ case, depending on the presence or absence of cosmological constant $\Lambda$, we see two different asymptotic late time behavior depicted in  Figs.~\ref{Fig:stiffpositive-Lambda},~\ref{Fig:stiffpositive-no-Lambda}. In particular, as Fig.~\ref{Fig:stiffpositive-no-Lambda} shows for $\Lambda=0$ case $\kappa, K$ remain finite in asymptotic future and the Universe do not isotropize. This shows the importance of $\Lambda$ in isotropization of the Universe, in accord with Wald's cosmic no-hair theorem \cite{Wald:1983ky}. 

\subsection{\texorpdfstring{$r= 0$}{} case, whimper singularity}\label{sec:whimper} 

In this case $\rho=\rho_0=const.$ remains finite at the BB. Nonetheless,  equations imply  $f<0$ for any generic $w$, i.e. we can't have both an almost constant energy density and a vanishing tilt. $f<0$ implies $\tanh\beta=1$ at the BB and we have a highly relativistic fluid whose velocity vector is null. Thus for the special case of radiation matter with $w=1/3$ the fluid curves and  the tilt direction are the same \cite{Madsen:1986mi}.
For this case, shear and $\beta$ should both be positive, $\mathfrak{S}=+1$. Unlike $r<0$ and generic $w$ case which is shear dominated, for $r=0$ case curvature, matter and shear are all of the same order and
\begin{equation}
    H\simeq \sinh\beta \left(\frac{A_0}{C}\rho_0^{\frac{w}{1+w}}+ \frac{C}{6A_0}(1+w) \rho_0^{\frac{1}{1+w}} \right)=H_0 \frac{\sinh\beta}{\sinh\beta_0},
\end{equation}
and
\begin{equation}\begin{split}
       h=s-3=f=-1-2K& , \qquad K=\frac{\sigma_0}{3H_0}= 1-\frac{A_0}{H_0}=1-\kappa,\qquad e^b=a^K,\\
   \rho_0\sinh^2\beta_0&=\frac{6 K A_0 H_0}{1+w}=\frac{6 K A_0^2}{(1+w)(1-K)},\qquad C=\rho_0^{\frac{w}{1+w}} \sinh\beta_0,
\end{split}
\end{equation}
That is,
\begin{equation}
    \Omega_\sigma=K^2,\qquad \Omega_k=(1-K)^2,\qquad \Omega_w=2K(1-K),
\end{equation}
only depend on $K$ and not $w$ or other parameters. Since the evolution of the system is only sensitive to $K$ and not $w$, we have plotted evolution of the Universe for a single value of $w$, $w=1/3$ in Fig.~\ref{Fig:w=1/3whimper}. Our numeric analyses (not shown here)  confirm that other $w$ show the same evolution.
\begin{figure}[!ht]
	\begin{center}
		\includegraphics[width=7.5cm]{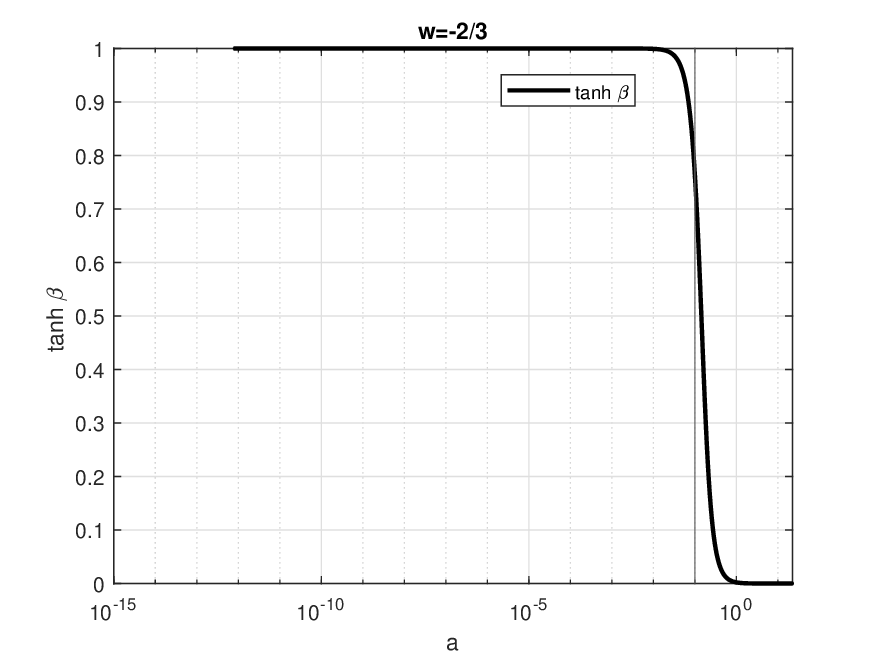}	
        \includegraphics[width=7.5cm]{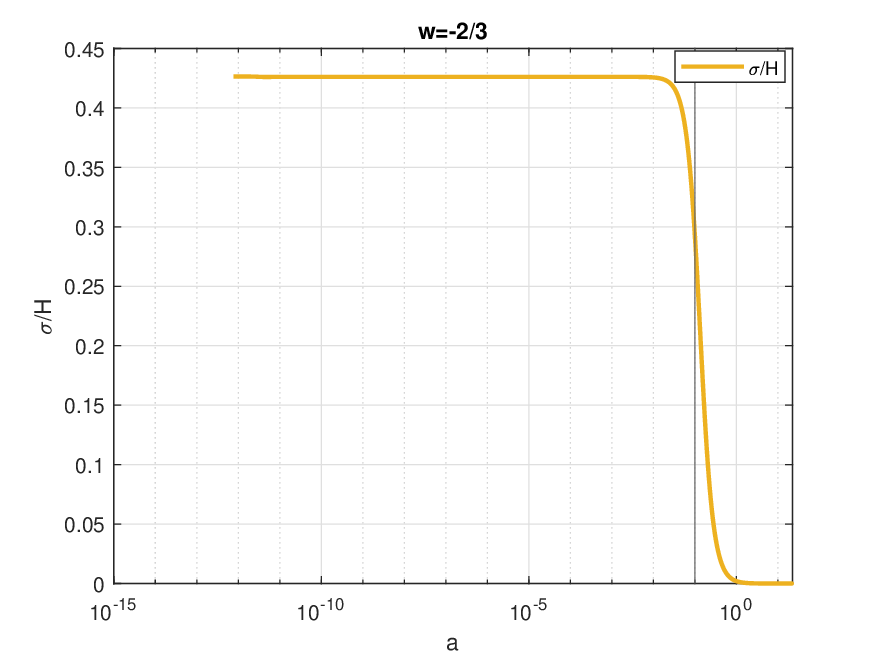}
        \includegraphics[width=7.5cm]{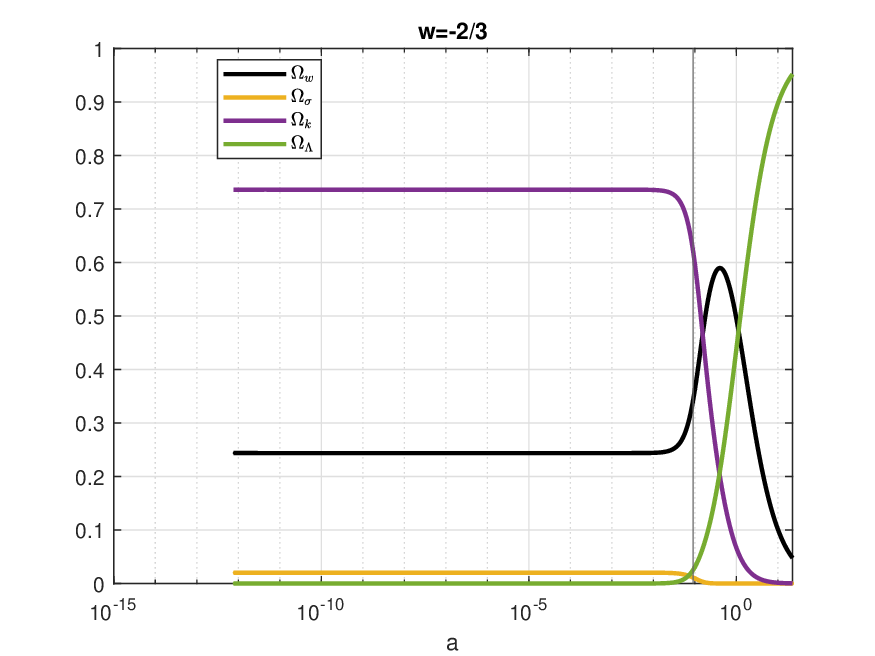}	
        \includegraphics[width=7.5cm]{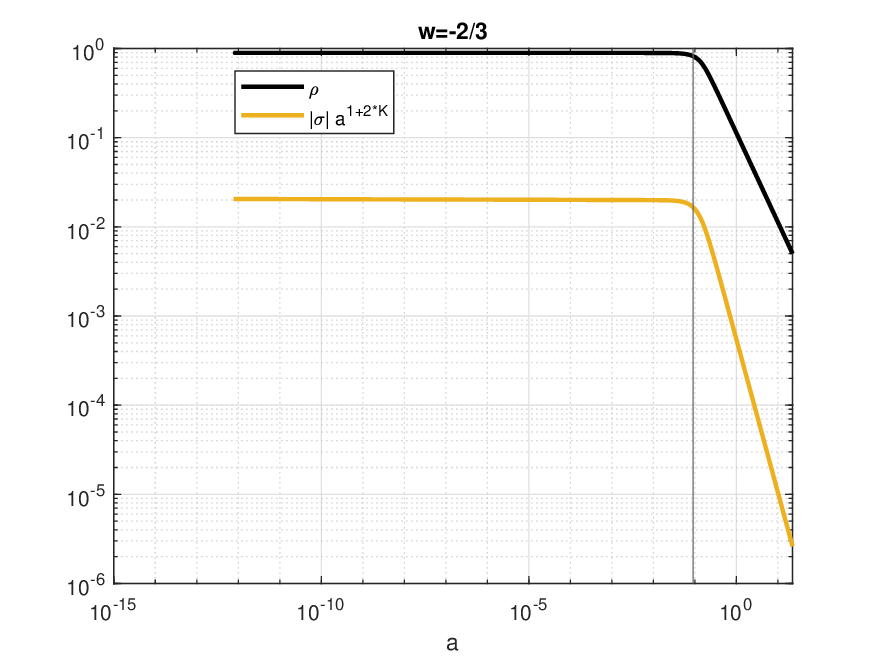}	
\caption{Whimper singularity case $r=0$ for $w=-2/3$. As discussed, for $r=0$ case, necessarily $\beta>0$. We we see near the BB, $\sinh\beta\propto a^f$ with $f<0$ and $\sigma/H$ can remain large before a quick drop  to zero around $a\sim 1$. In this case $\Omega_w, \Omega_k, \Omega_\sigma$ are generically of the same order and none of them do not generically dominate at the BB. They remain essentially constant up until the late Universe.}
		\label{Fig:w=-2/3whimper}
	\end{center}
\end{figure}

In terms of the cosmic comoving time, 
\begin{equation}
    a(t)= [(1+2 K)H_0 t]^{\frac{1}{(1+2 K)}}.
\end{equation}
The metric for this case is of the form
\begin{equation}\label{nearBB-metric-rho-const}
\begin{split}
    ds^2&=-dt^2+a^{4K+2}dz^2+a^{2-2K} e^{-2 A_0 z}(dx^2+dy^2)\\
    &=-dt^2+t^{2}dz^2+t^{\frac{2-2K}{1+2 K}} e^{-2 A_0 z}(dx^2+dy^2) .
\end{split}
\end{equation}
Since $0<K<1$, the $z$ direction shrinks while $x,y$ directions are large in the early $t\to 0$ Universe. That is, we have a \textit{pancake type Universe}. This is to be contrasted with the filament type Universe of the previous subsection. 

\paragraph{Singularity structure.} Since density $\rho$ and pressure $p$ remain finite and regular around $a\to 0$, Ricci scalar and all invariants built from powers of Ricci tensor remain finite. Moreover,  Kretschmann invariant \eqref{Kretschmann} for this case remains finite at $a\approx 0$. 
On the other hand, one can see that the electric part of the Weyl tensor \eqref{Electric-Weyl} is  $\mathcal{E}^x_x\approx\frac{1}{2}H^2 K (1+2K+f)\approx0$. Therefore, all invariants made from Riemann are finite (non-divergent) for this case. This may be understood noting that the source of divergence in components of Riemann is  coming from the tilt $\beta$ which is an observer (frame) dependent quantity and therefore scalars made out of Riemann are expected not to depend on $\beta$ and expected to remain finite. 

\begin{figure}[!ht]
	\begin{center}
		\includegraphics[width=7.5cm]{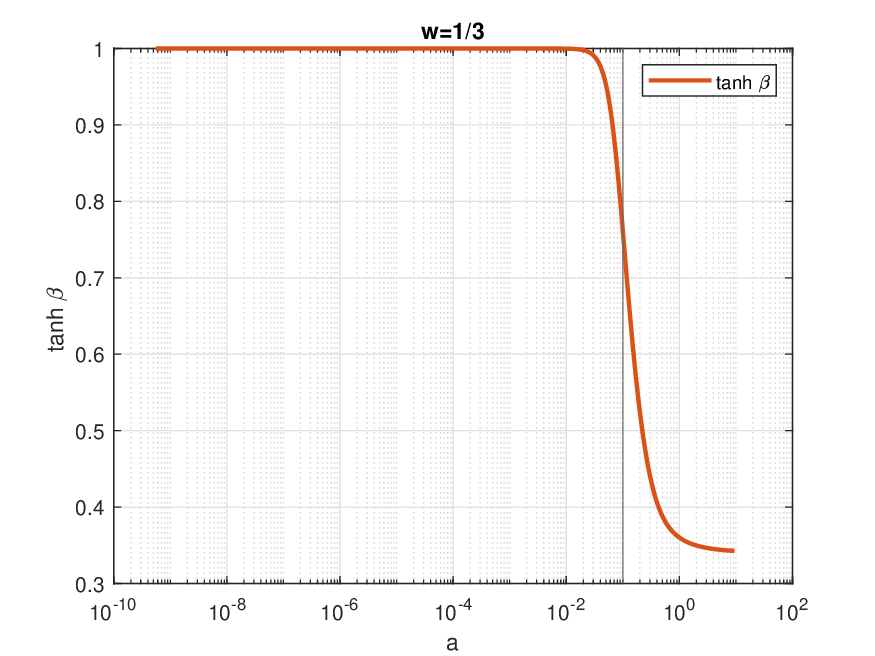}	
        \includegraphics[width=7.5cm]{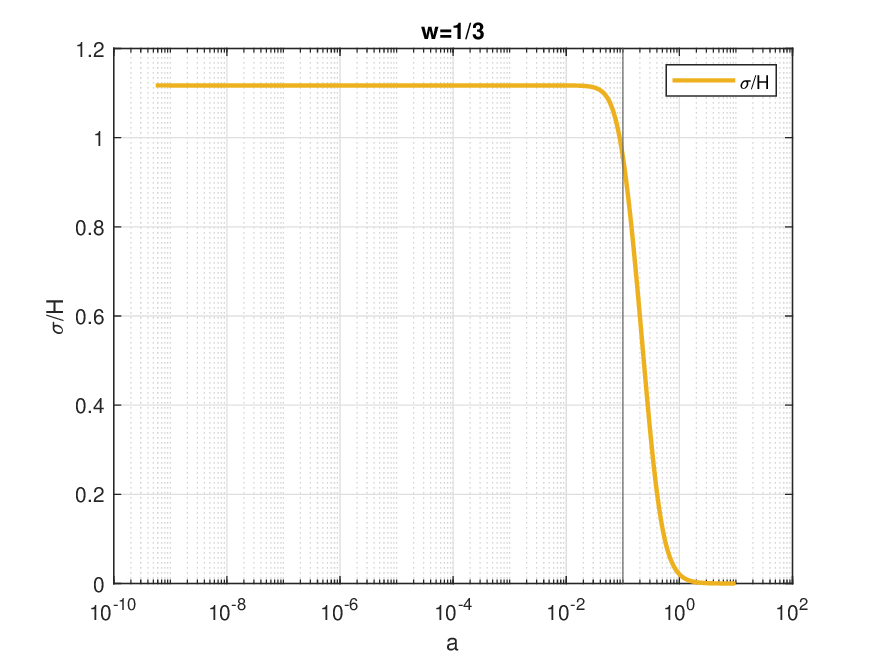}
        \includegraphics[width=7.5cm]{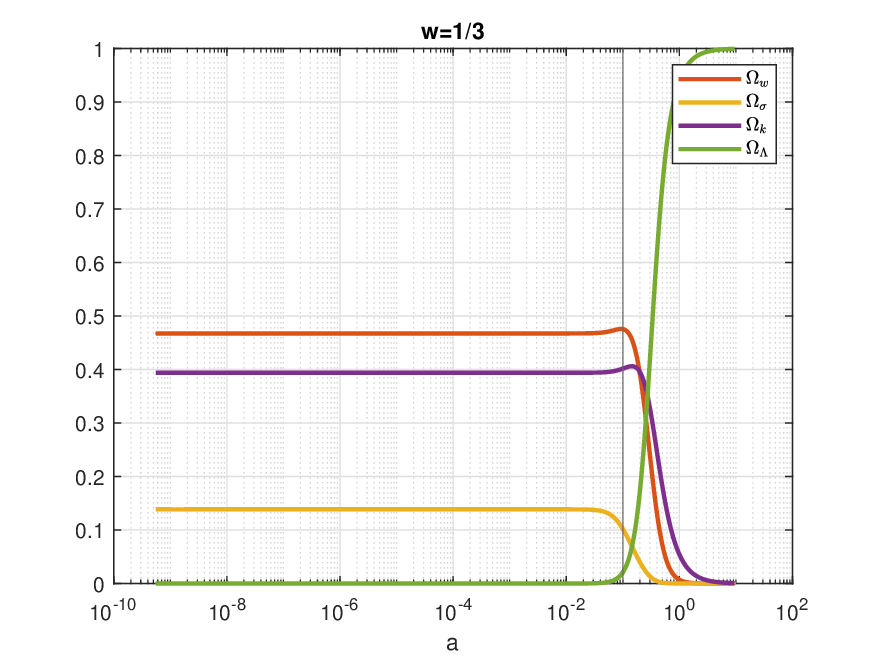}	
        \includegraphics[width=7.5cm]{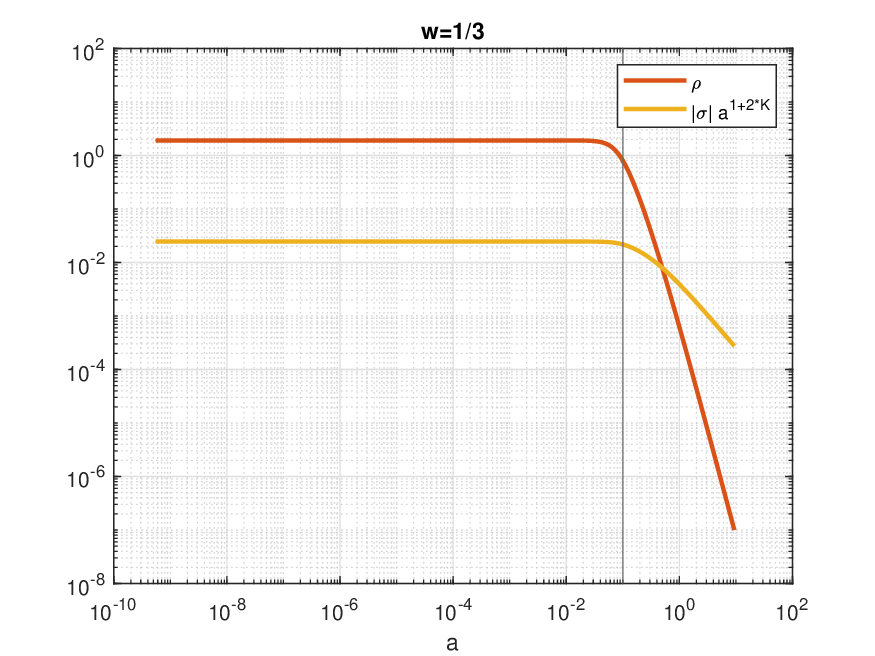}	
\caption{Whimper singularity case of $r=0$ for $w=1/3$. This case and $w=-2/3$ of Fig.~\ref{Fig:w=-2/3whimper} are qualitatively very similar.}
		\label{Fig:w=1/3whimper}
	\end{center}
\end{figure}

Singularities can hence appear only in the non-scalar part of the curvature. This kind of singularity has been called ``whimper singularity'' \cite{Ellis:1974ug}. (Ref.~\cite{Lim:2006iw} calls it ``kinematic singularity''.) The whimper singularity will hence show up in the geodesic deviation equation, through Riemann curvature \eqref{Riemann-frame}, which up to leading order is given as
\begin{subequations}\label{Riemann-frame-whimper}
\begin{align}
  &\Tilde{R}_{tyty}=\Tilde{R}_{txtx}= -H^2 \left(h-hK+ (K-1)^2\right)=3K(1-K)H^2,\\
 & \Tilde{R}_{tztz}= -H^2 \left(h+2hK+ (2K+1)^2\right)=0,\\
 & \Tilde{R}_{tyyz}=\Tilde{R}_{txxz}=3  K (1-K) H^2,\\
 & \Tilde{R}_{yzyz}=\Tilde{R}_{zxzx}= H^2\left(1-\kappa^2+K-2K^2\right)=3K(1-K)H^2,\\
 & \Tilde{R}_{xyxy}= H^2\left(1+K^2-\kappa^2-2K\right)=0. 
 \end{align}
\end{subequations}
The nonzero components diverge like $H^2$ at  the singularity. The particles feel infinite tidal forces near the singularity along $x$ and $y$ directions, while along the dipole direction $z$  tidal forces are finite. This supports the pancake Universe picture discussed above. 

To gain a better intuition for whimper singularity let us explore Raychaudhuri equation for the fluid velocity vector field $u^\mu$ \cite{EMM-2012}:
\begin{align}
    u^{\mu}\partial_{\mu}\theta+\frac{1}{3} \theta^2+ \sigma_{\mu\nu}\sigma^{\mu\nu}-\nabla_\mu a^\mu+\frac{1}{2} (\rho+ 3 p)=0
\end{align}
where ${\theta}=\nabla_\mu u^\mu$ is the expansion,  $a_\mu$ is its acceleration and $\sigma_{\mu\nu}=\frac12 h_\mu^\alpha h_\nu^\beta(\nabla_\alpha u_\beta+\nabla_\beta u_\alpha)-\frac13 \theta (g_{\mu\nu}+u_\mu u_\nu)$ and $h_\nu^\mu=\delta_\nu^\mu+u_\nu u^\mu$. When $r=0$, \eqref{Con-rho-i} and \eqref{Con-p-i} imply that the fluid does not expand and has zero acceleration, $\theta=0$  and $a^\mu=0$. Moreover, for our case
 third, fourth and fifth terms stay finite. In this way curvature singularity is avoided. 

In the table below we have summarized the discussions in this section about the singularity structure. 
\begin{table}
    \centering
    \begin{tabular}{|c|c|c|}\hline 
        $r<0, -1<w<1$ & $H=\frac{1}{3t}\sim a^{-3}$, filament Universe, shear and Weyl dominated singularity\\ \hline
         $r<0, w=1$& $H\sim a^{-3}$, Weyl-Ricci balanced singularity, open Kasner Universe\\ \hline
$r\simeq 0$ & $H\sim const.$, finite curvature invariants, whimper singularity, pancake Universe\\ \hline
    \end{tabular}
    \caption{{Big bang singularity structure for a single fluid dipole cosmology model. In all these cases Universe starts with large deviations from isotropy (with large shear), but as discussed, the shear dies off quite abruptly and metric isotropizes while the tilt $\beta$ can remain finite and can even grow \cite{Krishnan:2022uar}}.}
    \label{tab:Sec-3-singl.}
\end{table}

\section{Big Bang in Dipole \texorpdfstring{$\Lambda$CDM}{}}\label{sec4:Dipole-LCDM}

In this section we analyze the near BB behavior of dipole $\Lambda$CDM model which involves two fluids with $w=1/3$ and $w=0$, as well as the cosmological constant $\Lambda$ \cite{Ebrahimian:2023svi}.  While our analytic discussions are for generic initial values, for our numerical analysis we take the Planck $\Lambda$CDM values \cite{Planck-2018}.  This in particular means that in the units we have adopted here we evolve our Universe from $a=1$ to smaller values (and also bigger $a$ values) with the values of parameters $b(a=1)=0, \rho_m(a=1)= 0.9, \rho_r(a=1)=3\times 10^{-4}, \Lambda=2.1$ \cite{Ebrahimian:2023svi}. {These values are in our adapted units which are equivalent to the dimensionless quantities  $\Omega_{\Lambda}\approx 0.7 ,\Omega_m\approx 0.3,\Omega_{r}\approx 10^{-4}$ at $a=1$. These values are very close to the Planck $\Lambda$CDM values for the cosmological parameters.} Tilts and $A_0$ have no Planck $\Lambda$CDM counterparts, for $A_0$ we choose $A_0=0.1$ which is compatible with Planck bound on the spatial curvature. Values $\beta_r(a=1), \beta_m(a=1)$ are different for each plots and we have quoted the corresponding values in the figure captions. 

\subsection{General analysis}\label{sec:LCDM-general}

This is a special 2 fluid ($w=0, 1/3$) case of the general multiple fluid cosmology. The system has $2+4$ degrees of freedom, $a,b$ (or $H,\sigma$) and $\rho_r, \beta_r; \rho_m, \beta_m$. Equations governing these variables in terms of a function of scale factor $a$ take the following form 
\begin{subequations}\label{EEQLCDM}
\begin{align}
     &f_m:=\frac{d\ln\sinh\beta_m}{d\ln a}=-1-2\mathfrak{S} K,\\
&f_r:=\frac{d\ln\sinh\beta_r}{d\ln a}=-\frac{2}{3-\tanh^{2}\beta_{r}}[3\mathfrak{S} K+\kappa \tanh\beta_r],\\
      &r_m:= \frac{d\ln\rho_m}{d\ln a}=  -[(1-\tanh^{2}{\beta_{m}})+2(1-\mathfrak{S} K\tanh^{2}{\beta_m}-\kappa \tanh{\beta_{m}})],\\
  &r_r:= \frac{d\ln\rho_r}{d\ln a}=  -\frac{4}{3-\tanh^{2}\beta_{r}}[(1-\tanh^{2}{\beta_{r}})+2(1-\mathfrak{S} K\tanh^{2}{\beta_r}-\kappa \tanh{\beta_{r}})],\\
&s:=3+\frac{d\ln\sigma}{d\ln a}=\frac{\mathfrak{S}}{K}[\Omega_m\tanh^2\beta_m +\frac{4\tanh^2\beta_r}{3+\tanh^2\beta_r}\Omega_r],\\
&{h:=\frac{d\ln H}{d\ln a}=-2+2\Omega_\Lambda-\Omega_\sigma+\Omega_k +\frac{1}{2\cosh^2\beta_m}\Omega_m,}
\end{align}
\end{subequations}
where $K, \mathfrak{S}, \kappa$ are defined as in \eqref{kappa-K-S} and \eqref{Omega-def} and imply
\begin{equation}\label{S-r-m-sign}
\begin{split}
    &\Omega_\Lambda+\Omega_\sigma+\Omega_k +\Omega_m+\Omega_r=1,\qquad \Omega_\sigma=K^2,\ \Omega_k=\kappa^2\\
    &  2K\kappa \mathfrak{S}=\Omega_m \tanh\beta_m+ \Omega_r \frac{4\tanh\beta_r}{3+\tanh^2\beta_r},
\end{split}
\end{equation}
with
\begin{equation}
\Omega_m=\frac{\rho_m}{3H^2}\cosh^2\beta_m, \qquad \Omega_r=\frac{\rho_r}{3H^2}\cosh^2\beta_r (1+\frac13\tanh^2\beta_r).
  \end{equation}

From the above we learn,
\begin{equation}\label{K-s-m-r}
    \begin{split}
        K\mathfrak{S} (s-2\kappa \tanh\beta_r) &= \Omega_m \tanh\beta_m (\tanh\beta_m-\tanh\beta_r),\\
        K\mathfrak{S} (s-2\kappa \tanh \beta_m) &= \Omega_r \frac{4\tanh\beta_r}{3+\tanh^2\beta_r} (\tanh\beta_r-\tanh\beta_m).
    \end{split}
\end{equation}
We also note that,
\begin{equation}
\begin{split}
 2\kappa K=|\Omega_m \tanh\beta_m&+ \Omega_r \frac{4\tanh\beta_r}{3+\tanh^2\beta_r}|<\Omega_m+\Omega_r<1,\\
&(\kappa+\mathfrak{S}K)^2<1.
\end{split}    
\end{equation}
The above then implies $r_m, r_r\leq 0$ and $h<0$ for the whole course of evolution, regardless of the initial values and the sign of $\beta_r,\beta_m, \sigma$. Moreover, $r_m\simeq 0$ ($r_r\simeq 0$) can only happen when $\beta_m$ ($\beta_r$) becomes very large so that $|\tanh\beta_m|\simeq 1$ ($|\tanh\beta_r|\simeq 1$) and also when $\kappa+\mathfrak{S} K=1$. For these to happen we need $\beta_m, \beta_r>0$ (which implies $\mathfrak{S}=+1$). This extends our results of section \ref{sec:whimper} to the dipole $\Lambda$CDM case, where the Universe can start with a whimper singularity.

{One can  integrate  continuity equations for pressureless matter and radiation sectors to get (see \cite{Ebrahimian:2023svi} for more details)} 
\begin{equation}\label{PLCDM}
\begin{split}
ae^{2b}\sinh{\beta_{m}} = C_{m}\\
\rho_{r}^{1/4}\ ae^{2b}\sinh{\beta_{r}} = C_{r}
\end{split}
\end{equation}
Note that $\beta_m, \beta_r$ (and hence $C_m, C_r$) can be positive or negative. The sign of shear $\mathfrak{S}$ is positive (negative) if $\beta_r, \beta_m$ are both positive (negative). When $\beta_r, \beta_m$ have different signs, $\mathfrak{S}$ can have different signs and importantly as our numeric analysis also shows, $\mathfrak{S}$ can change in the course of evolution of the Universe. However, as discussed below \eqref{Hdot-sigmadot} and \eqref{kappa-K-S}, $\mathfrak{S}$ can only evolve from $-1$ to $+1$, i.e. if we start from $\mathfrak{S}=+1$, there is no chance of change of sign for it, while for $-1$ initial value we may end up with a Universe with $\mathfrak{S}=+1$. The same argument also shows that the change of sign in the shear can happen at most only once.

\subsection{Near BB expansion}\label{sec:near-BB-LCDM}

While the above equations are true for any $a$, one may focus on the near BB $a\to 0$ behavior of the system. The analysis here are direct extensions of what we had  in the single fluid case, so we will be brief here and do not repeat all the equations. As before, it is the sign of $\beta_r, \beta_m$ and $\mathfrak{S}$ which determine the near BB behavior. As \eqref{S-r-m-sign} indicates, 
\begin{equation}\label{sign-S-Omega}
    \mathfrak{S}=\text{sgn}(\Omega_m \tanh\beta_m+ \Omega_r \frac{4\tanh\beta_r}{3+\tanh^2\beta_r}).
\end{equation}
Signs of $\beta_r, \beta_m$ can't change in the course of evolution, whereas $\mathfrak{S}$ can change sign \cite{Ebrahimian:2023svi}. Thus, there are 4 different cases depending on signs of $\beta_r, \beta_m$, which we analyze separately. 

\subsubsection{\texorpdfstring{${\beta_{r} < 0}$ and  ${\beta_{m} <0}$}{} case}\label{sec:betar-ve-betam-ve}

For this case $\mathfrak{S}=-1$ and it always remains negative in the evolution, $r_m, r_r<0$ and
\begin{equation}\label{f--ve-ve}
    f_m=2K-1,\qquad {f_r=\frac{3}{3-\tanh^2\beta_r}(2K+\frac23 \kappa |\tanh\beta_r|)}, \qquad r_r=-4-4f_r+8K.
\end{equation}
$f_r>0$ and $\beta_r$ goes to zero near the BB and $r_r=-4$. The BB is shear dominated which will be followed by a radiation dominated era, i.e. near BB we have $K=1$ and $\sigma=-3H$ and $H=H_0 a^{-3}$. $f_m$ can take different signs depending on $K$. So, near BB $f_m=1, f_r=2$ implying $|\sinh\beta_m|\sim a, |\sinh\beta_r|\sim a^2$; both $|\beta_r|, |\beta_m|$ grow and $|\beta_r|$  grows faster. 

We have a curvature singularity near $a=0$, $R\sim \rho_r\sim a^{-4}$ whereas electric Weyl components grow like $H^2\sim a^{-6}$. That is, the singularity is Weyl dominated. Metric near the BB has the same form as in \eqref{nearBB-metric-generic} and we have a filament-type Universe. 

\begin{figure}[!ht]
	\begin{center}
\includegraphics[width=7.5cm]{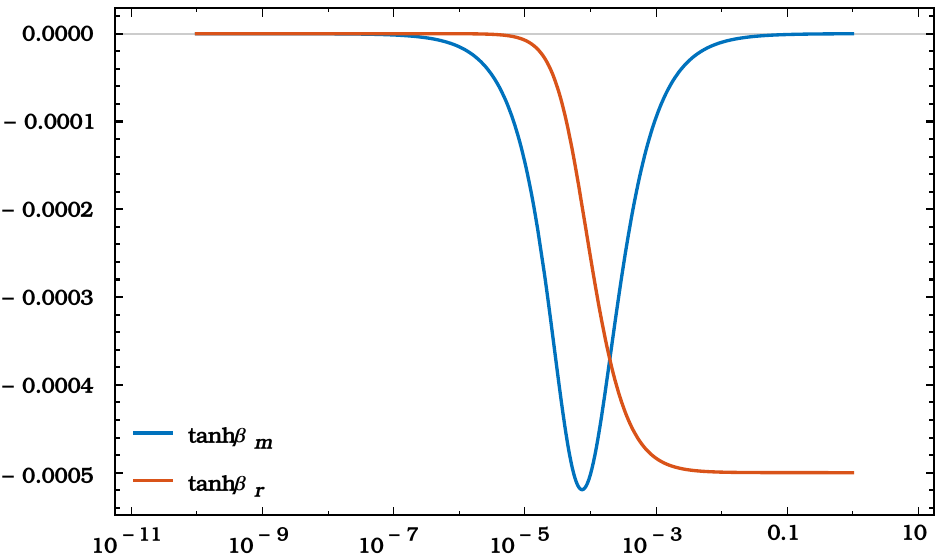}
 \includegraphics[width=7.5cm]{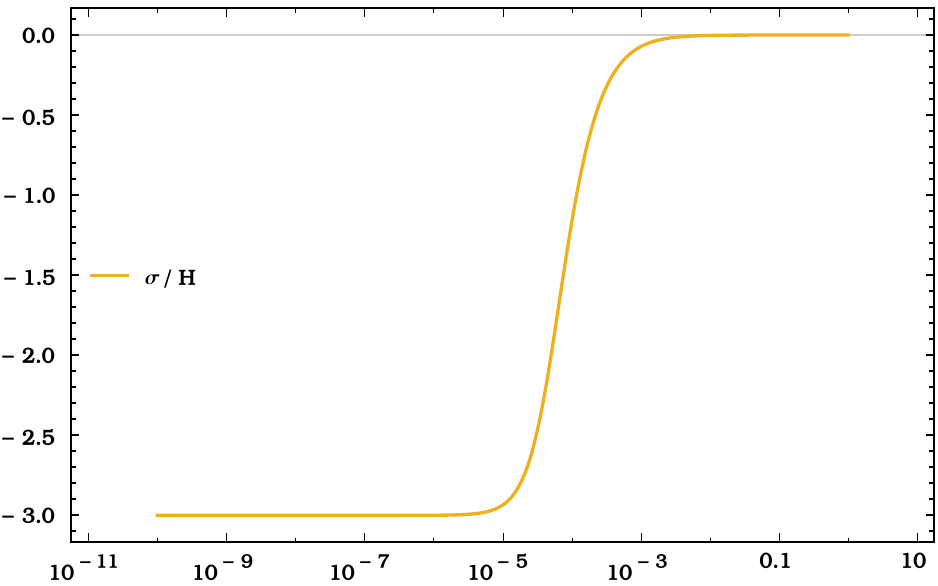}	
        \includegraphics[width=7.5cm]{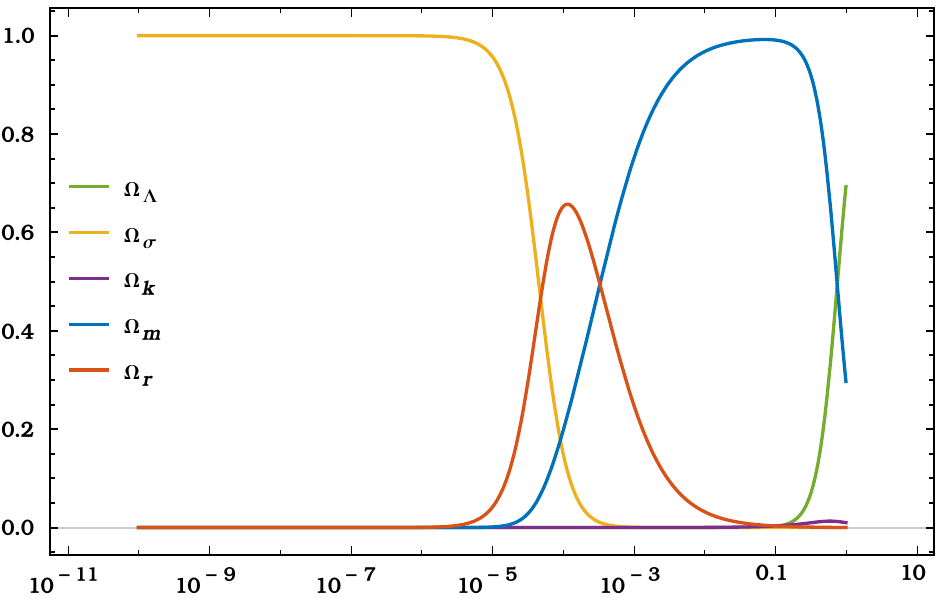}	
         \includegraphics[width=7.5cm]{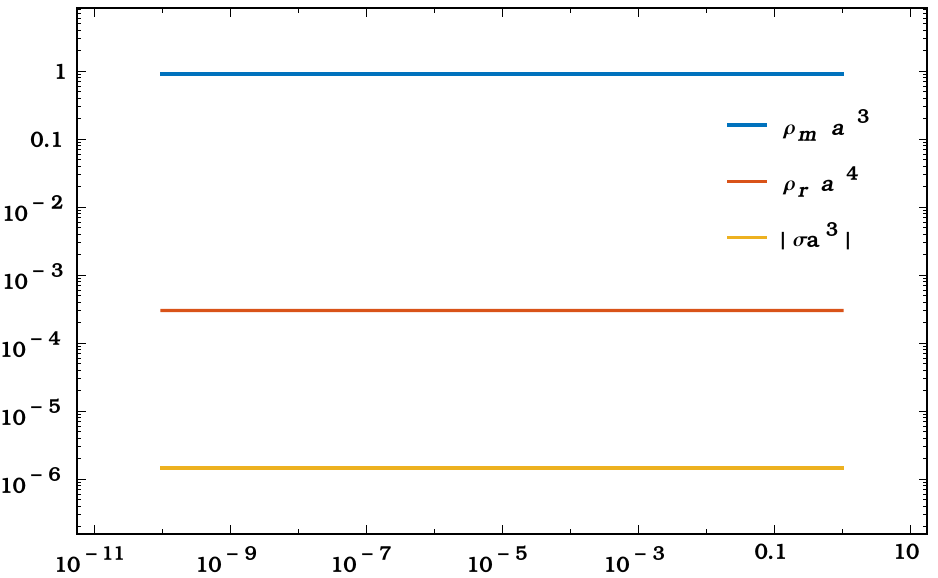}	     	
       \caption{Plots of $\tanh{\beta_r},\tanh{\beta_m}$, $\sigma/H$, $\Omega_{i}$s, $\rho_{m}, \rho_r$ for $\beta_r,\beta_m<0 $ case discussed in section \ref{sec:betar-ve-betam-ve} for boundary conditions  $\beta_{r}(1) = -5 \times 10^{-4}$, $\beta_{m}(1) = -10^{-7}$. $\mathfrak{S}$ approaches to $-1$ towards $a \rightarrow 0$. While $\rho_r a^4,\rho_m a^3$ go to a constant near the BB, $\sigma\sim a^{-3}$ and hence the BB is shear-dominated and the curvature singularity is Weyl-dominated.}\label{Fig:SN-rN-mN}
	\end{center}
\end{figure}

As we evolve to larger $a$ values, shear drops to zero, Universe isotropizes fast and $K$ drops to its minimum $K=0$. $f_m$ then changes sign and hence $|\beta_m|$ is expected to reach a maximum value, before going to zero at late times. Nonetheless, $|\beta_r|$ continues its mild, but monotonic grows. These are depicted in Fig.~\ref{Fig:SN-rN-mN}.

\subsubsection{\texorpdfstring{${\beta_{r} > 0}$ and  ${\beta_{m} > 0}$}{} case}\label{sec:betar+ve-betam+ve}

\begin{figure}[!ht]
	\begin{center}
\includegraphics[width=7.5cm]{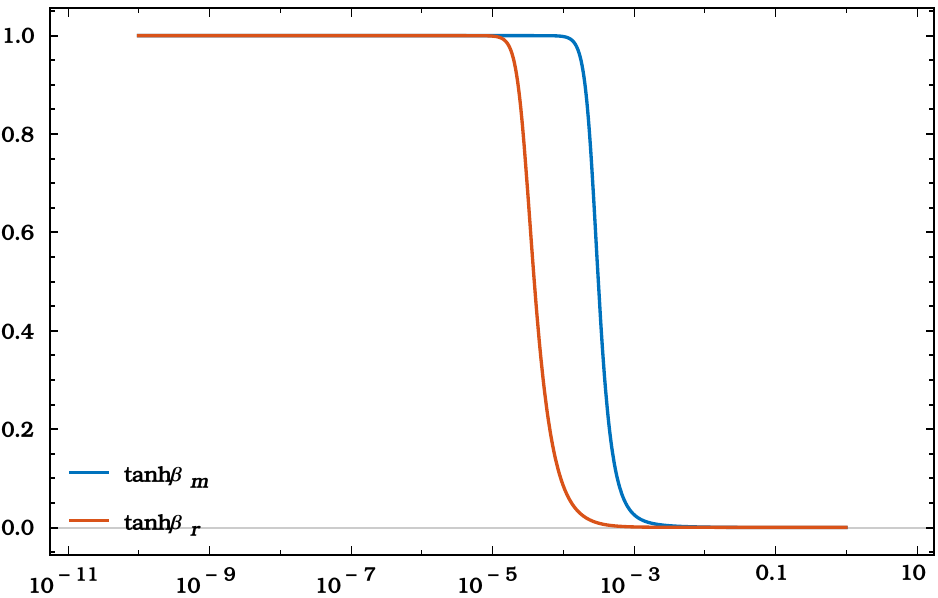}
 \includegraphics[width=7.5cm]{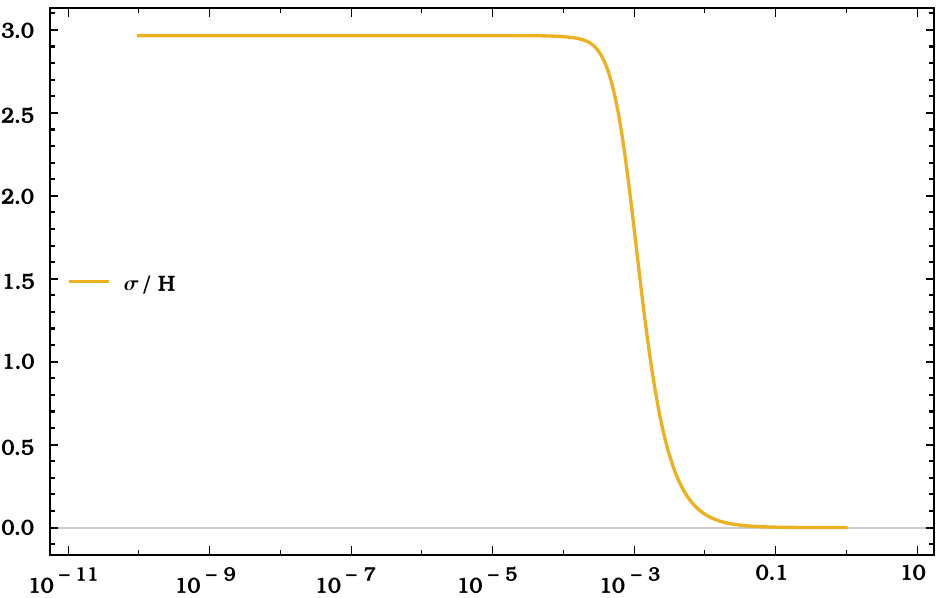}	
        \includegraphics[width=7.5cm]{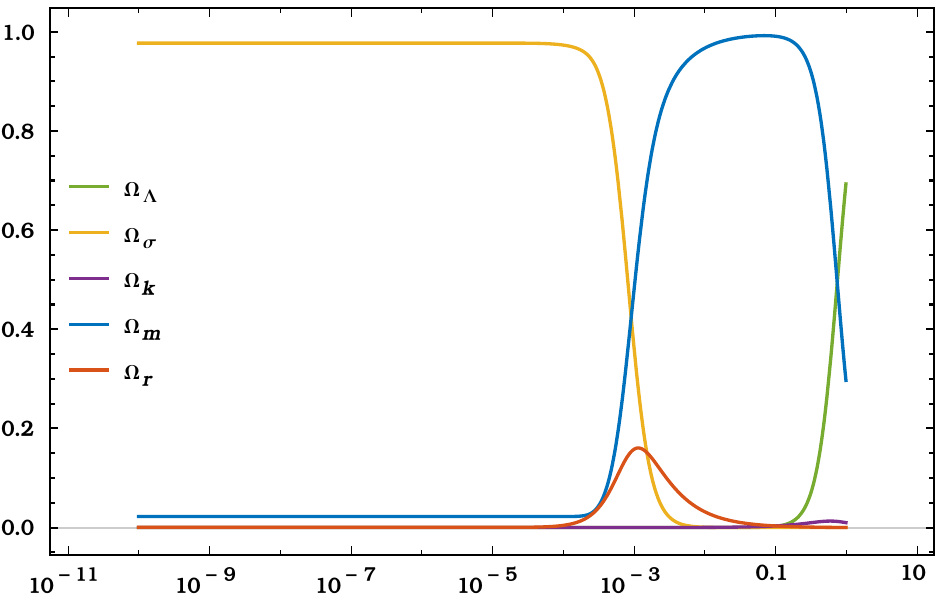}	
         \includegraphics[width=7.5cm]{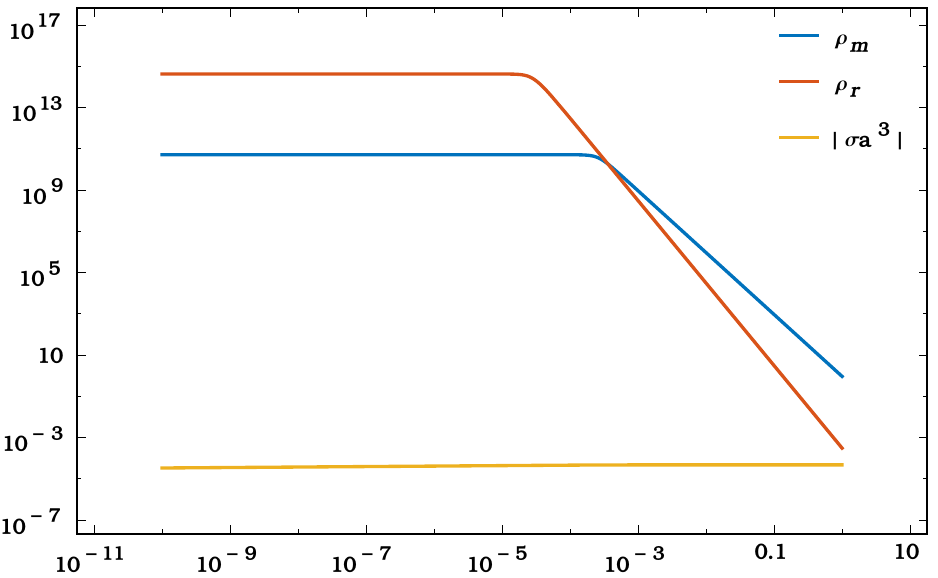}	     	
       \caption{Plots of $\tanh{\beta_r},\tanh{\beta_m}$, $\sigma/H$, $\Omega_{i}$s, $\rho_{m}, \rho_r$ for $\beta_r,\beta_m>0 $ case discussed in section \ref{sec:betar+ve-betam+ve} for boundary conditions  $\beta_{r}(1) = 5 \times 10^{-4}$, $\beta_{m}(1) = 10^{-5}$. For all the cases $\mathfrak{S}$ approaches to $+1$ towards $a \rightarrow 0$, $\rho_r,\rho_m$ go to a constant near the BB and we have a whimper singularity.}\label{Fig:SP-rP-mP}
	\end{center}
\end{figure}

For this case $\mathfrak{S}=+1$ and always remains positive in the evolution, and
\begin{equation}\label{f-+ve+ve}
    f_m=-2K-1,\qquad f_r=-\frac{2}{3-\tanh^2\beta_r}(3K+ \kappa \tanh\beta_r).
\end{equation}
$f_r, f_m<0$ hence $\beta_r,\beta_m$ become large  near the BB ($\tanh\beta_r\simeq \tanh\beta_m\simeq 1$), $2K\kappa\simeq \Omega_m+\Omega_r$ and $s\simeq 2\kappa$. Noting that $\Omega_\Lambda\ll 1$, then, $\kappa+K=1$ and hence $r_r\simeq 2 r_m\simeq -4(1-K-\kappa)\simeq 0$. So, for this case $\rho_m, \rho_r$ remain finite and  $\Omega_\sigma, \Omega_k, \Omega_m, \Omega_r$ can be of the same order near the BB. This implies there is no curvature singularity. This is another example of whimper singularity, where all curvature invariants are finite and singularity only appears in components of Riemann tensor (in a tetrad frame) and in geodesic deviation equation, cf. discussions in section \ref{sec:whimper}. As we evolve to larger $a$ values, shear drops to zero, Universe isotropizes fast and the tilts  drop to zero. These are depicted in Fig.~\ref{Fig:SP-rP-mP}.

\subsubsection{\texorpdfstring{${\beta_{r}> 0}$ and  ${\beta_{m} < 0}$}{} case}\label{sec:betam-ve-betar+ve}

In this case $\mathfrak{S}$ can take either signs near the BB and can change sign in the course of evolution of the Universe. So, we consider the two $\mathfrak{S}=+1,-1$ cases for near the BB, separately. As we  argue below we can't have $\mathfrak{S}=+1$ arbitrarily close to the BB.

\paragraph{$\mathfrak{S}=+1$ is not possible arbitrarily close to the BB.} We prove this by contradiction. Assuming that this statement is not true, for $\mathfrak{S}=+1$ we have
\begin{equation}
    f_m=-2K-1,\qquad f_r=-\frac{2}{3-\tanh^2\beta_r}(3 K+ \kappa \tanh\beta_r).
\end{equation}
Since $f_m, f_r<0$ both tilts are large near the BB, $\tanh\beta_r\simeq 1, \tanh\beta_m\simeq -1$ and hence,
\begin{equation}
\begin{split}
    r_m\simeq -2(1-K+\kappa),&\qquad r_r\simeq -4(1-K-\kappa)\\
    2K\kappa=\Omega_r-\Omega_m,\qquad f_r =&-(3K+\kappa), \qquad K s=\Omega_m+\Omega_r
\end{split}\end{equation}
The above implies $r_m=\frac12 r_r-4\kappa<0$ and  $r_r\leq 0$. Recalling \eqref{sign-S-Omega}, $\mathfrak{S}=+1$ is possible only when $\Omega_r\geq\Omega_m$. On the other hand, $\Omega_r/\Omega_m\sim a^{4\kappa}$. However, no matter how small $\kappa$ is, there exists $a\sim 10^{-N}$ such that $4\kappa N>1$ and hence $\Omega_r\lesssim \Omega_m$. That is, $\mathfrak{S}$ can't remain positive for large enough $N$ (very close to the BB). 

The above result may be described in a different way. Let us suppose that at late times we have an almost isotropic Universe with $K\to 0^+$ and evolve it backwards. Regardless of the details of values of $\Omega_m, \Omega_r; \beta_m, \beta_r$ at late times (e.g. at $a=1$), evolving back the system in time, one will reach to an (almost) shear dominated Universe with $K=1$ ($\mathfrak{S}=+1$). If we continue evolving further back in time, closer to the BB, as the above argument shows, one would eventually cross to $\mathfrak{S}=-1$, $K=1$ regime at the BB. 

\begin{figure}[!ht]
	\begin{center}
\includegraphics[width=7.5cm]{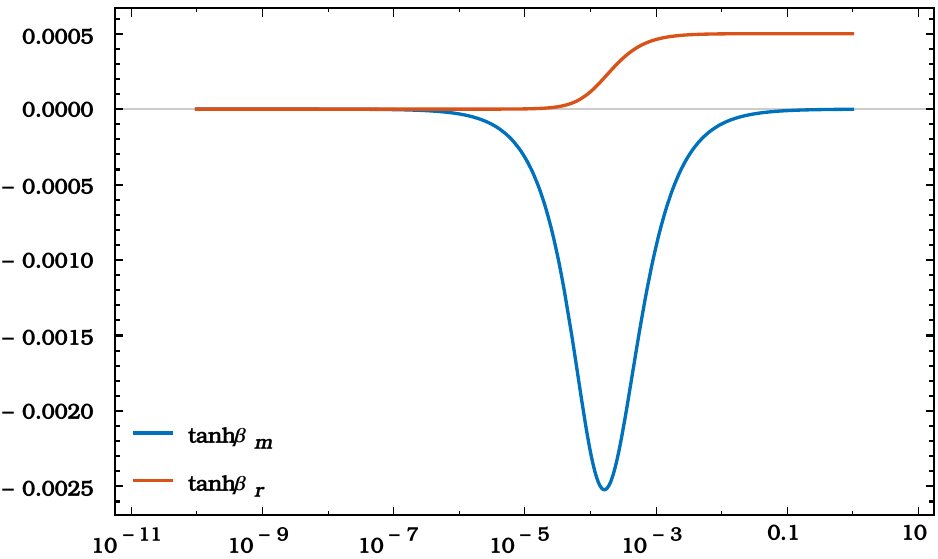}	
        \includegraphics[width=7.5cm]{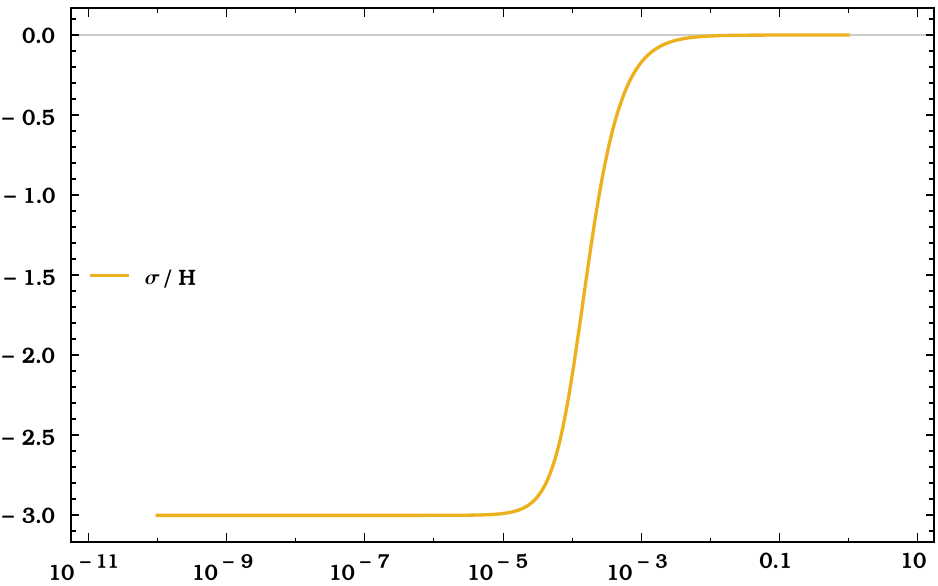}	
         \includegraphics[width=7.5cm]{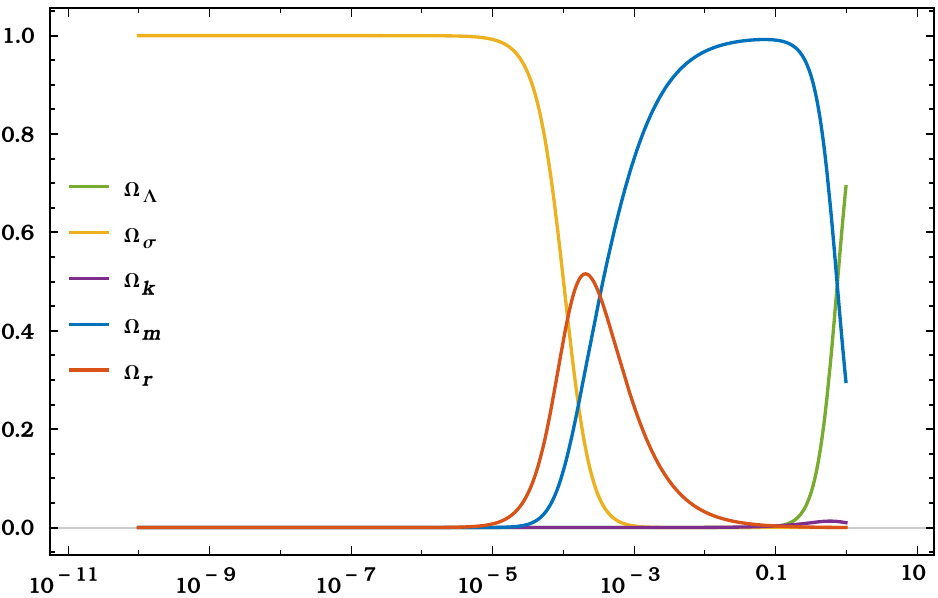}	
        \includegraphics[width=7.5cm]{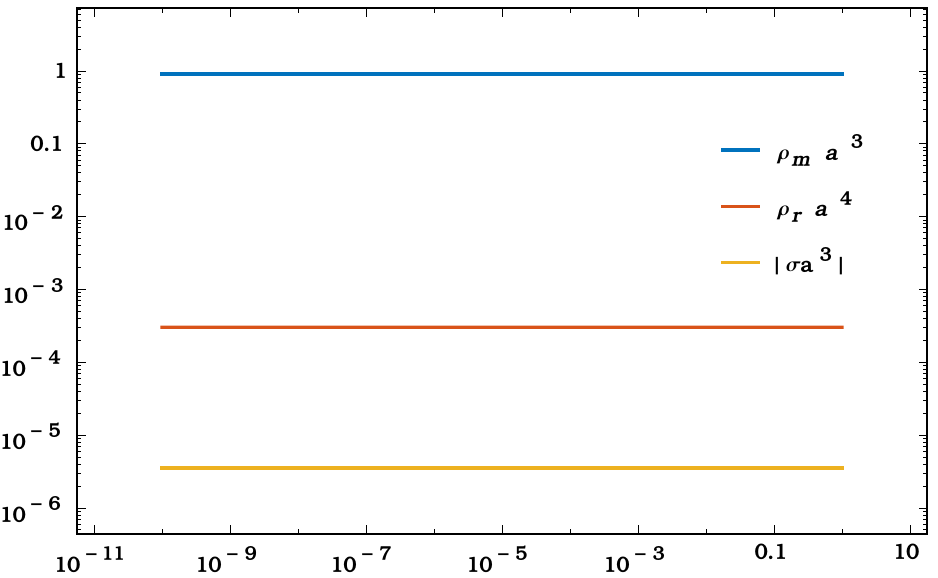}	
       \caption{$\tanh{\beta_r}, \tanh{\beta_m}$, $\sigma a^{3}$,  $\Omega_{i}$s and $\rho_{r}, \rho_m$ for $\beta_r>0, \beta_m<0, \mathfrak{S}=-1$ (cf.  section \ref{sec:betam-ve-betar+ve})  with boundary conditions $\beta_{m}(1) = -10^{-6}$, $\beta_{r}(1) = 5 \times 10^{-4}$. In this case $\sigma$ does not change sign. }\label{Fig:SN-rP-mN-SN}
	\end{center}
\end{figure}

\paragraph{$\mathfrak{S}=-1$.} For this case,
\begin{equation}
\begin{split}
    &f_m=2K-1,\qquad f_r=-\frac{2}{3-\tanh^2\beta_r}(-3 K+ \kappa \tanh\beta_r),\\
          &r_m=  -2 -[\frac{1}{\cosh^{2}{\beta_{m}}}+2(K\tanh^{2}{\beta_m}+\kappa |\tanh{\beta_{m}|})]<-2,\\     
  &r_r =  -4-\frac{8}{3-\tanh^{2}\beta_{r}}(K\tanh^{2}{\beta_r}-\kappa \tanh{\beta_{r}}),\\
\end{split}
\end{equation}
where $ae^{2b}= a^{1-2K}$ and $\Omega_k/\Omega_r\sim a^{-r_m}$. Since $r_m<0$ and $\Omega_k, \Omega_m$ must remain between 0 and 1, $\kappa\simeq 0$ and hence $f_r=6K/(3-\tanh^2\beta_r)>0$ . So, $\tanh\beta_r\simeq 0$ and $r_r=-4, f_r=2K$. Moreover, from \eqref{K-s-m-r} we learn that $\Omega_r=|\tanh\beta_m| \Omega_m$. In this case $r_m, r_r<0$ and hence we have curvature singularity. Equations for $s, h$ imply that we should be dealing with a shear dominated regime with $K=1$ and $\sigma=-3H, H=H_0 a^{-3}$, so both tilts become small at the BB and the curvature singularity is Weyl dominated. 

\begin{figure}[!ht]
	\begin{center}
\includegraphics[width=7.5cm]{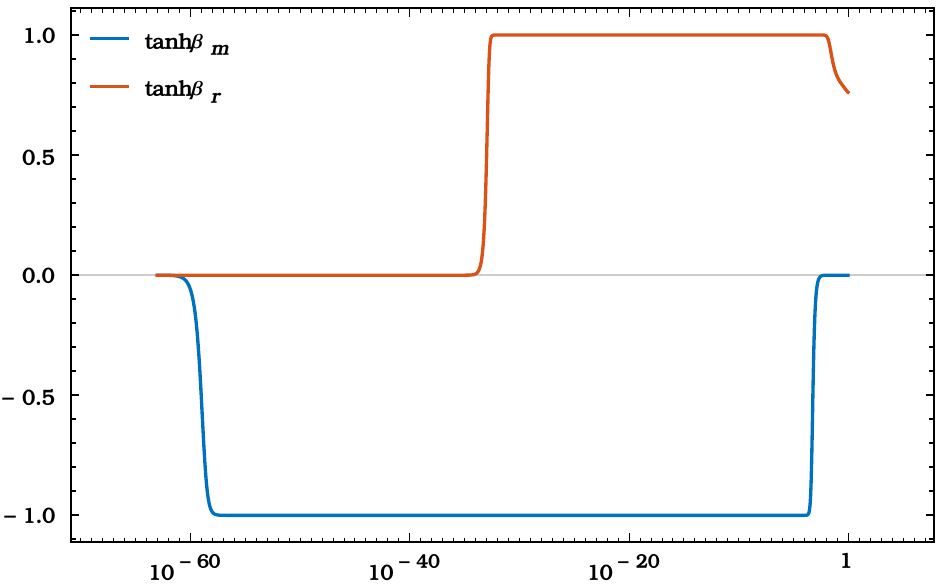}	
        \includegraphics[width=7.5cm]{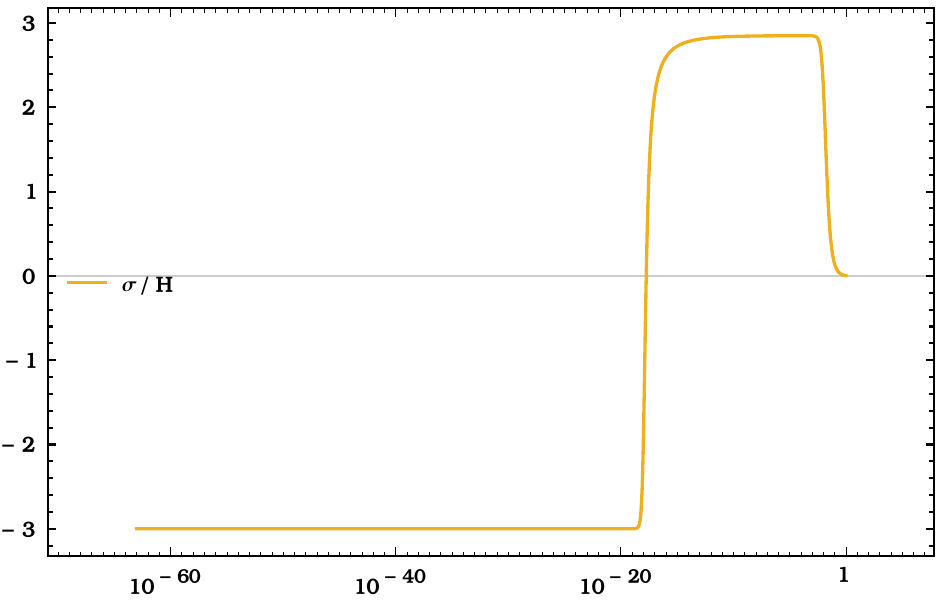}	
         \includegraphics[width=7.5cm]{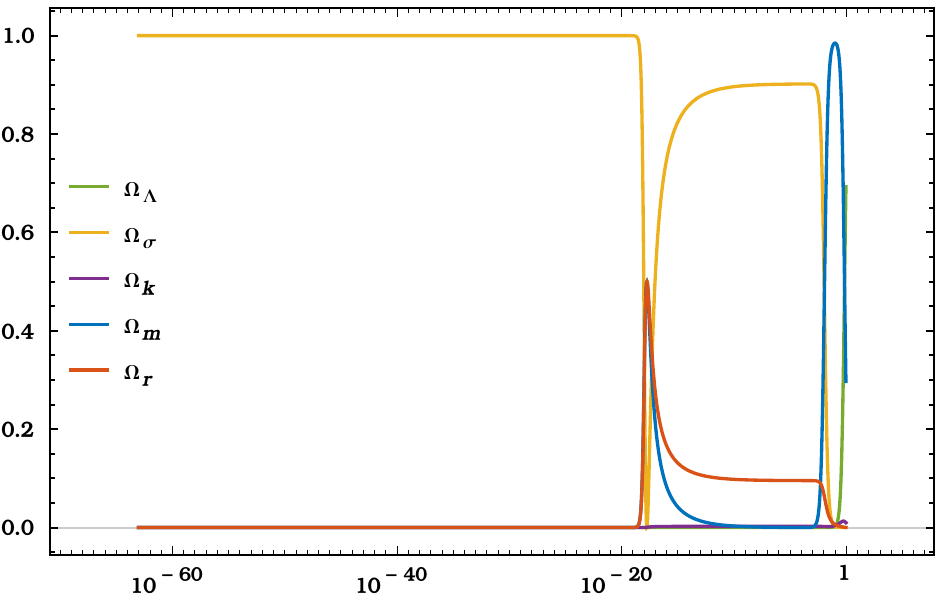}	
        \includegraphics[width=7.5cm]{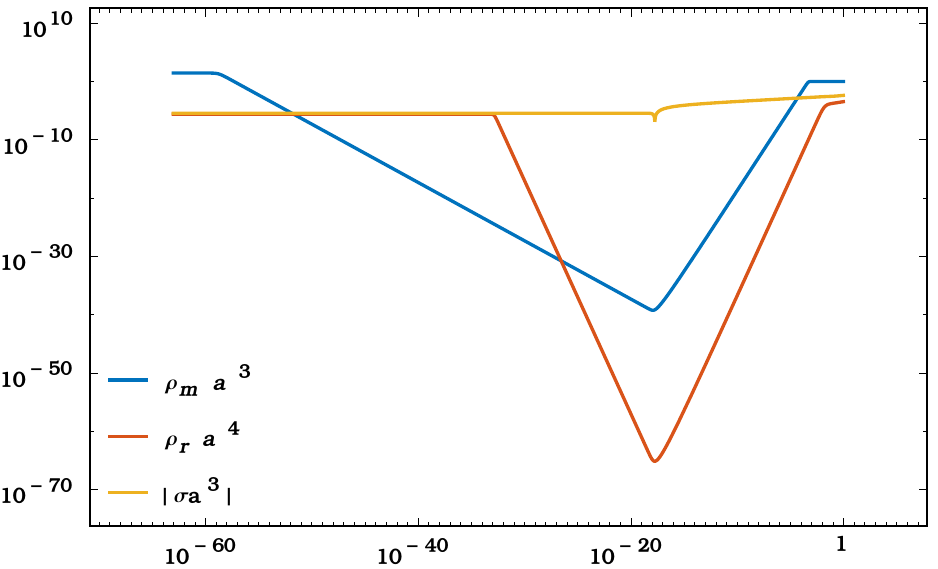}	
       \caption{$\tanh{\beta_r}, \tanh{\beta_m}$, $\sigma /H$,  $\Omega_{i}$s and $\rho_{r}, \rho_m$ for $\beta_r>0$ and $\beta_m<0$ (cf. section \ref{sec:betam-ve-betar+ve})  for boundary conditions  $\beta_{m}(1) = -5\times10^{-7}$, $\beta_{r}(1) =1$. $\mathfrak{S}$ changes sign from $-1$ to $+1$ at an intermediate time which depends on the boundary values.}\label{Fig:SN-rP-mN-SP}
	\end{center}
\end{figure}

As we evolve to larger $a$ values, shears drops to zero, Universe isotropizes, i.e. $K$ drops to zero. $f_m$ can change sign and $|\beta_m|$ has hence a maximum value.  For the case,  depending on the boundary values we have two possibilities: There is no change of sign in $\mathfrak{S}$ and $\mathfrak{S}K$ goes from $-1$ to $0^-$ in asymptotic future. Or there is a change of sign, $K\mathfrak{S}$ first changes values from $-1$ to $+1$ fast, stays on for few e-folds  and then drops to zero, $K\mathfrak{S}$ evolves as $-1\to +1\to 0^+$.   These are depicted in Figs.~\ref{Fig:SN-rP-mN-SN},  \ref{Fig:SN-rP-mN-SP}. For the case in Fig.~\ref{Fig:SN-rP-mN-SN} $\beta_r$ goes to a non-zero constant value at late times, while $\beta_m$ drops to zero \cite{Ebrahimian:2023svi}.

\subsubsection{\texorpdfstring{${\beta_{r}<0}$ and ${\beta_{m}>0}$}{} case}\label{sec:betar-ve-betam+ve}
\begin{figure}[!ht]
	\begin{center}
 \includegraphics[width=7.5cm]{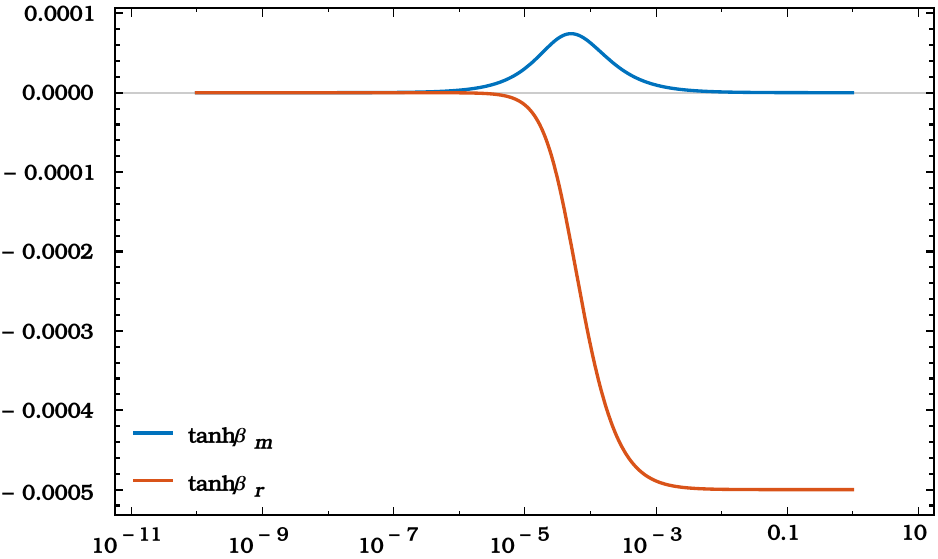}	
        \includegraphics[width=7.5cm]{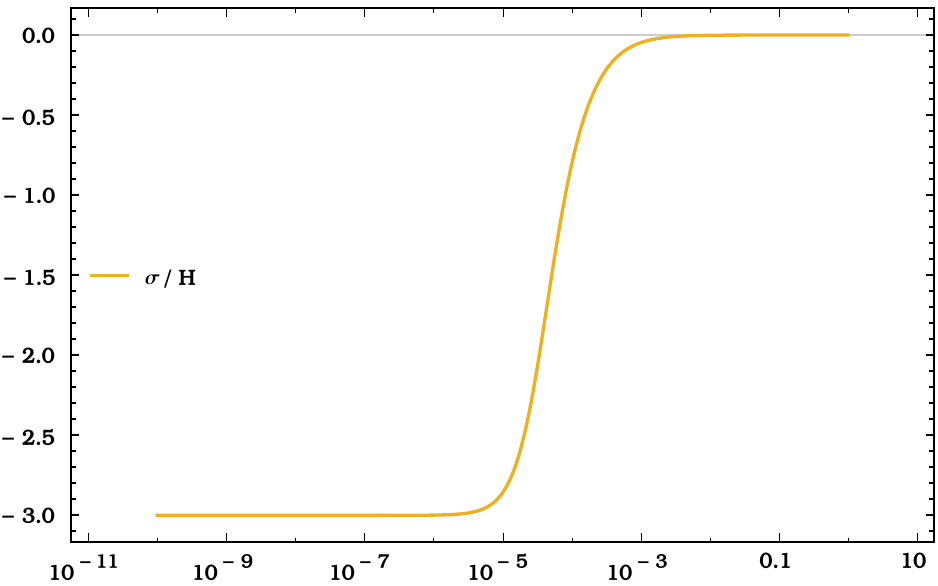}	
         \includegraphics[width=7.5cm]{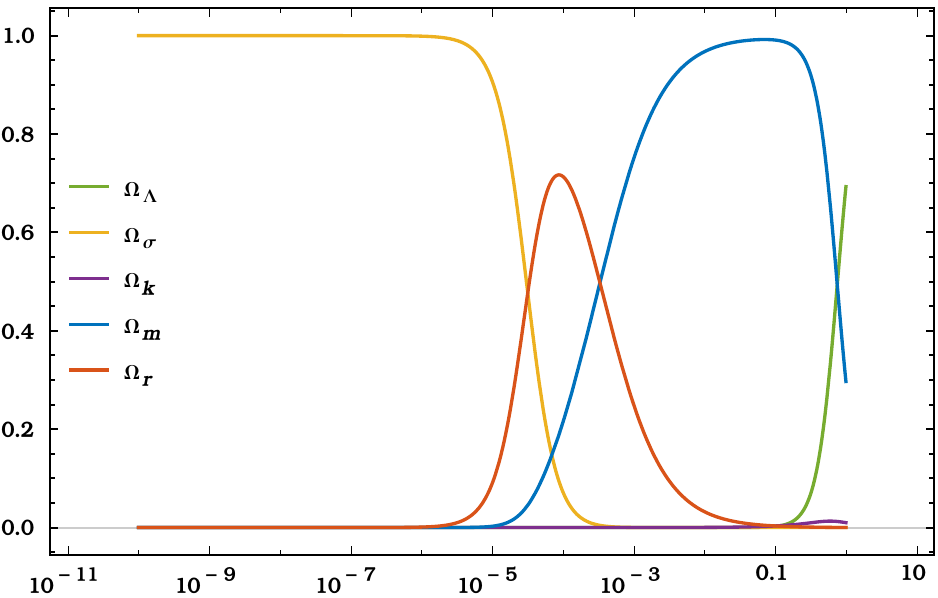}	
        \includegraphics[width=7.5cm]{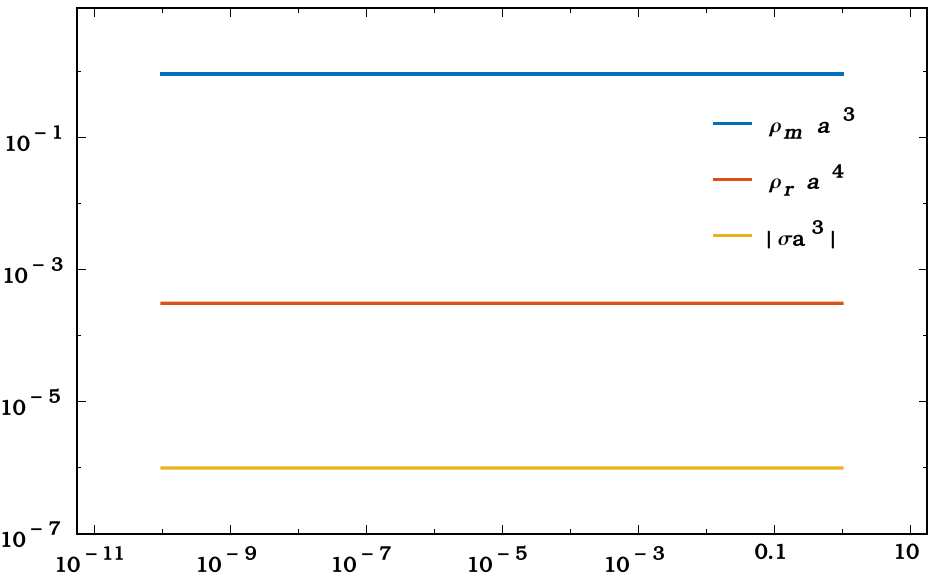}	
       \caption{Plots of $\tanh{\beta_r}, \tanh{\beta_m}$, $\sigma/H$,  $\Omega_{i}$s and $\rho_{r}, \rho_m$ for  $\beta_r<0, \beta_m>0$ and $\mathfrak{S}=-1$ (discussed in section \ref{sec:betar-ve-betam+ve}),  for boundary conditions  $\beta_{m}(1) = 10^{-8}$, $\beta_{r}(1) =- 5 \times 10^{-4}$. Note that in the course of evolution $\mathfrak{S}$ does not change sign and remains $-1$. }\label{Fig:SN-rN-mP-SN}
	\end{center}
\end{figure}

\begin{figure}[!ht]
	\begin{center}
\includegraphics[width=7.5cm]{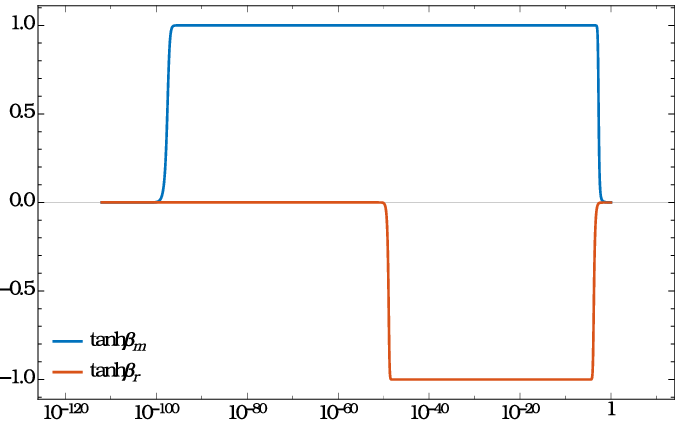}	
        \includegraphics[width=7.5cm]{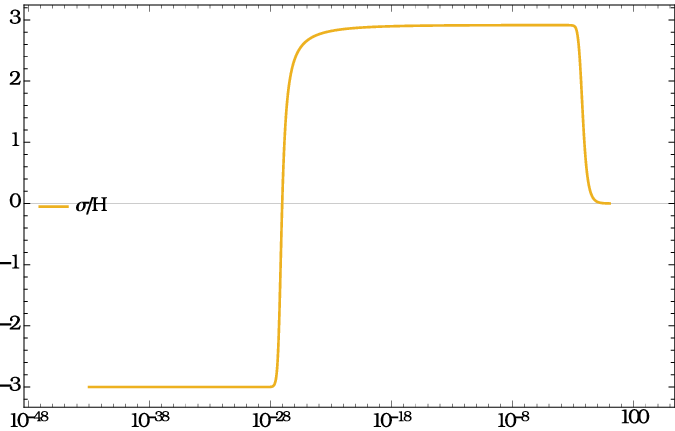}	
         \includegraphics[width=7.5cm]{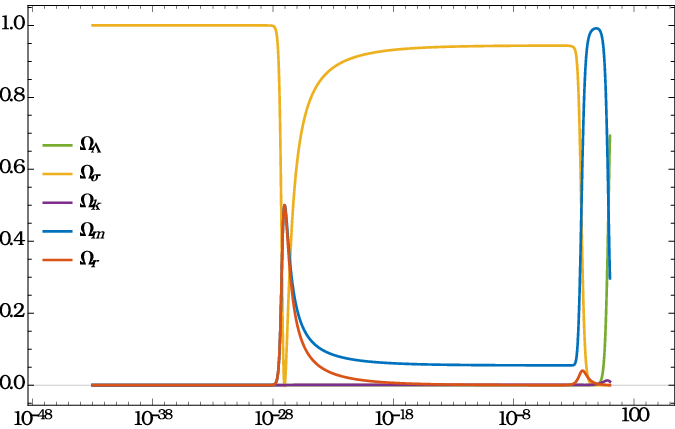}	
        \includegraphics[width=7.5cm]{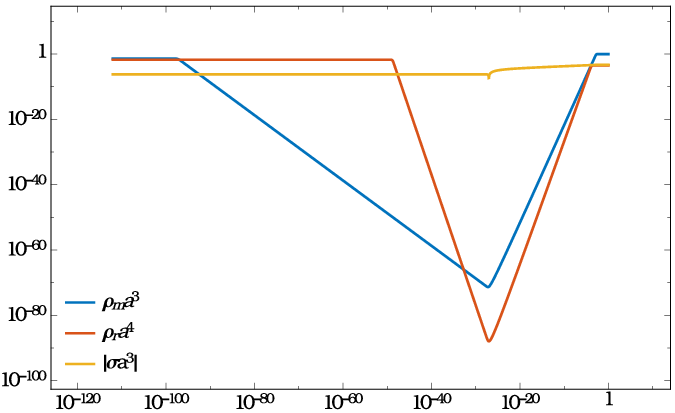}	
       \caption{$\tanh{\beta_r}, \tanh{\beta_m}$, $\sigma a^{3}$,  $\Omega_{i}$s and $\rho_{r}, \rho_m$ plots for $\beta_r<0$ and $\beta_m>0$ (discussed in section \ref{sec:betar-ve-betam+ve}). $\mathfrak{S}$ changes sign from $-1$ to $+1$ at an intermediate time for boundary conditions  $\beta_{m}(1) =10^{-4}$, $\beta_{r}(1) =-5\times 10^{-4}$.   $\Omega_{i}$s are plotted in logarithmic scale and hence the sharp in $\Omega_{\sigma}$ just shows the change of sign in $\sigma$. }\label{Fig:SN-rN-mP-SP}
	\end{center}
\end{figure}

The analysis of this case is similar to the previous case discussed in section \ref{sec:betam-ve-betar+ve}. The $\mathfrak{S}=+1$ at the BB is not possible and we have only $\mathfrak{S}=-1$ near the BB, which is a Weyl dominated, curvature singularity with $\sigma=-3H$ with vanishing tilts at the BB. For this case too, we have possibility of change of sign for which $K\mathfrak{S}$ evolves as $-1\to +1\to 0^+$ and the case without change of sign where $K\mathfrak{S}$ evolves as $-1\to 0^-$.   The evolution for this two cases is depicted in Figs.~\ref{Fig:SN-rN-mP-SN} and \ref{Fig:SN-rN-mP-SP}. 

\section{Discussion and outlook}\label{sec5:closing}

In this work we focused on the near Big Bang (BB) behavior in dipole cosmology models. Our analysis has revealed that depending on the sign of the shear near the BB, there are two general behaviours: For negative shear, curvature invariants blow up, in particular the dominant singularity is in the Weyl-squared tensor, which is dominated by the shear. While the Ricci scalar (or Ricci-squared) is also divergent, they are subdominant to the Weyl invariant. 
For this case tilt remains small near the BB and as the metric \eqref{nearBB-metric-generic} indicates,  near the BB one essentially finds a $1+1$ ``filament type'' Universe. For positive shear, however, we find that while the curvature invariants are all finite, the tilt blows up and we have a whimper singularity \cite{Ellis:1974ug}. That is, the matter is erupting at the speed of light in a cone with axis along the $z$ direction. The whimper singularity shows itself in the geodesic deviation equation. It would be instructive to study further near whimper singularity for these cases and whether it can have further cosmological implications for early Universe cosmology and inflation. Moreover, the cosmologies we have studied in section \ref{sec3:near-BB-general-w} for $\Lambda=0$ cases can have future singularity, as discussed in \cite{Lim:2006iw}. It is also desirable to analyze future $(a>1)$ behaviour of the system especially when the relative tilts can grow indefinitely in the asymptotic future. 

In our analysis  in section \ref{sec3:near-BB-general-w} we limited ourselves to fluids which respect dominant energy condition (DEC).  We saw that the cases respecting strong energy condition (SEC), $-1/3<w<1$ and those violating it $-1<w<-1/3$ show some different features: for the former tilt is vanishingly small near the BB and for the latter it blows up. Also, in section \ref{sec:r-ve-stiff-matter} we already found that the case of stiff matter, which is the borderline in DEC allowed fluids, it shows peculiar features not shared by $-1<w<1$ cases. One may explore the cases which violate DEC ($w>1$)  or even the phantom case which violate NEC,  e.g. as in \cite{Erickson:2003zm}.

In the dipole $\Lambda$CDM model we faced different cases depending on the signs of $\beta_r, \beta_m, \sigma$. Unlike the single fluid case the sign of shear, $\mathfrak{S}$, is not specified by the sign of the tilts; for the cases where tilts have the same sign, $\mathfrak{S}$ has also that sign and the sign does not change in the course of evolution. We have summarize all possible initial conditions in the BB in Table \ref{Nu}. 
For the cases where $\beta_m, \beta_r$ have different signs, $\mathfrak{S}$ can take either signs. Moreover, for $\mathfrak{S}=-1$ and for tilts of different signs, there is also possibility of $\mathfrak{S}=-1$ in the whole course of evolution or change to $\mathfrak{S}=+1$. So, all in all we have 6 different cases for the choices of the signs which are depicted in Figs.~\ref{Fig:SN-rN-mN}, \ref{Fig:SP-rP-mP},    \ref{Fig:SN-rP-mN-SN}, \ref{Fig:SN-rP-mN-SP},   \ref{Fig:SN-rN-mP-SN}, \ref{Fig:SN-rN-mP-SP}. In 1 of these 6 cases $\mathfrak{S}=+1$ near the BB depicted Fig.~\ref{Fig:SP-rP-mP}, $|\tanh\beta_m|\simeq |\tanh\beta_r|\simeq 1$. In the rest of 5 cases $\mathfrak{S}=-1$ and we have a shear dominated curvature singularity with vanishingly small $\beta_r, \beta_m$ near the BB; in 2 cases among the 5, cf. Fig.~\ref{Fig:SN-rP-mN-SP}, \ref{Fig:SN-rN-mP-SP}, $\mathfrak{S}$ can change sign from $-1$ to $+1$. As the figures show, in the intermediate $\mathfrak{S}K=+1$ phase $|\tanh\beta|=1$ for both radiation and dust sectors, i.e. the radiation and dust are moving at the speed of light but in opposite directions. 
\begin{table}
	\begin{center}
		\begin{tabular}{|c|c|c|c|c|}
			\hline
			$\mathfrak{S}K$&$-1$&$1$&$-1$&$-1$\\[0.1ex]
			\hline
			$v_m$&$0^{-}$&$1$ &$0^{-}$&$0^{+}$ \\
			\hline
			$v_r$&$0^{-}$&$1$ &$0^{+}$&$0^{-}$\\  \hline
			$Singularity$&$\tiny{Weyl}$&$whimper$ &$\tiny{Weyl}$&$\tiny{Weyl}$\\
			\hline	
           Depicted in  &Fig.~\ref{Fig:SN-rN-mN}& Fig.~\ref{Fig:SP-rP-mP}&Figs.~\ref{Fig:SN-rP-mN-SN},\ref{Fig:SN-rP-mN-SP}& Figs.~\ref{Fig:SN-rN-mP-SN},\ref{Fig:SN-rN-mP-SP}\\\hline
		\end{tabular}
		\caption{Summary of possible initial conditions near BB. We show sign of $\mathfrak{S}$ and  the corresponding velocities, $v_m=\tanh{\beta_m}$, $v_r=\tanh{\beta_r}$. $0^+$ ($0^-$) means the velocity approaches zero and is positive (negative). The two cases in each of the last two columns can be distinguished by the fact that one involve sign change in $\mathfrak{S}$ during the evolution and the other not.} 
		\label{Nu}
	\end{center}
\end{table}

Regardless of the behavior near the BB in all 6 cases of the dipole $\Lambda$CDM, at late times Universe isotropizes (fast) $\sigma$ goes to zero, $\beta_m$ also goes to zero while $\beta_r$ approaches  a finite asymptotic value. This asymptotic value depends on the initial conditions. That $\beta_m$ goes to zero and that the relative value of $\beta_m, \beta_r$ remains finite when the two have opposite signs, was already observed in \cite{Ebrahimian:2023svi}. However, our careful near BB analysis reveals that the same is also true for all the 6 cases. This may be traced to the fact that starting from the BB, the $\mathfrak{S}=+1$ case correspond to a whimper singularity for which $\beta_m, \beta_r$ are maximal, i.e. $|\tanh\beta_m|=|\tanh\beta_r|=1$. 

Starting from the asymptotic future  in the dipole $\Lambda$CDM and looking backward in time, we see that always $K\sim 0$, but in fact there are two possibilities, $K\sim 0$ with $\mathfrak{S}=+1$ or $-1$. The $K\mathfrak{S}\sim 0^-$ can only come from evolution of $\mathfrak{S}=-1, K\simeq 1$ case which has curvature singularity at the BB, cf. Figs.~\ref{Fig:SN-rN-mN},  \ref{Fig:SN-rP-mN-SN},   \ref{Fig:SN-rN-mP-SN}. The $K\mathfrak{S}\sim 0^+$ case, however, can be traced back to $\mathfrak{S}=+1, K\simeq 1$ near the BB, which is a whimper singularity, cf. Fig.~\ref{Fig:SP-rP-mP}, or to $\mathfrak{S}=-1, K\simeq 1$ at the BB with a change of sign in the shear in some intermediate time, cf. Figs.~\ref{Fig:SN-rP-mN-SP},  \ref{Fig:SN-rN-mP-SP}.

After this detailed study of dipole $\Lambda$CDM evolution, we are ready to study more closely the cosmography of this model and confront it directly with the cosmological data, especially those hinting to or reporting a dipole \cite{Ebrahimian-in-preparation}. {Moreover, here we focused on the background while to connect to real Universe one should also study anisotropic cosmic perturbation theory, e.g. see \cite{Notari:2015kla, Roldan:2016ayx}. These line are postponed to future investigations.}

\noindent\textbf{Acknowledgments}

We would like to thank Chethan Krishnan for discussions. AA, EE and MMShJ would like to acknowledge   Iran National Science Foundation (INSF) Grant No.  4026712. AA acknowledges the support of School of Astronomy of IPM. MMShJ is grateful to ICTP for the support from ICTP associates office (under Senior Associate program) and the HECAP section for hospitality. 

\appendix

\section{Energy conditions and the tilt}\label{appendix}

For completeness we briefly review energy conditions and how they appear in the tilted fluids. See \cite{Hawking:1973uf, Grumiller:2022qhx} for more discussions and references. Consider a prefect fluid the energy momentum tensor of which in some frame is given as ${\cal T}_{\mu\nu}$ such that in a comoving frame, 
${\cal T}^\mu{}_\nu=diag(-\rho, p, p, p)$. Also, we usually assume an Equation of State (EoS) $w$, $p=w \rho$. 

\paragraph{Null Energy Condition (NEC).} If for any future oriented null vector field $n^\mu$, ${\cal T}_{\mu\nu} n^\mu n^\nu\geq 0$ our fluid satisfies NEC. In the frame which is comoving with the fluid, this yields $\rho+p\geq 0$. Since by definition NEC has a generally invariant form, this condition should hold for our fluid in the tilted (non-comoving frame) too. Nonetheless, let us consider tilted energy momentum in~\eqref{T-uu}. For a generic null vector $n^\mu$, we get 
\begin{equation}    \label{NEC}
T_{\mu\nu}n^\mu n^\nu=(\rho+p) (n\cdot u)^2\geq 0
\end{equation}
where $u^\mu$ is the velocity vector field of the fluid \eqref{u-mu}. This implies $\rho+p\geq 0$ or $w\geq -1$.

\paragraph{Weak Energy Condition (WEC).} If for any timelike vector $t^\mu$, ${\cal T}_{\mu\nu} t^\mu t^\nu\geq 0$ our fluid satisfies WEC, i.e.
\begin{equation}\label{WEC}
T_{\mu\nu}t^\mu t^\nu=(\rho+p) (t\cdot u)^2+ p t^2\geq 0,\qquad t^2<0 
\end{equation}
Since $(t\cdot u)^2\geq |t^2|$, we learn that $\rho+p\geq 0, \rho\geq 0$. That is, $w\geq -1, \rho\geq 0$.

\paragraph{Strong Energy Condition (SEC).} SEC stipulates that for any timelike vector $t^\mu$, 
\begin{equation}\label{SEC}
(T_{\mu\nu}-\frac12 T g_{\mu\nu})t^\mu t^\nu=(\rho+p) (t\cdot u)^2+\frac12 (\rho-p) t^2\geq 0,\qquad T=T_{\mu\nu}g^{\mu\nu}
\end{equation}
Since $(t\cdot u)^2\geq |t^2|$, we learn that $\rho+p\geq 0, \rho+3p\geq 0$. That is, $w\geq -1/3$.

\paragraph{Dominant Energy Condition (DEC).} If for any future oriented causal vector field $c^\mu$, 
\begin{equation}\label{DEC}
P_\mu:= -T_{\mu\nu} c^\nu, \qquad P_\mu c^\mu \leq 0,
\end{equation}
and $P_\mu$ is a future oriented causal vector, the fluid respects DEC. For our tilted fluid, this implies,
\begin{equation}
P_\mu=(\rho+p) (c\cdot u) u_\mu+ p c_\mu=\text{future oriented causal vector},\qquad \rho+p\geq 0, \quad \rho>0,
\end{equation}
from which we learn that $\rho\geq |p|$. That is, $-1\leq w\leq 1$.  DEC implies speed of sound in the fluid remains subluminal and its violation leads to faster than light sound waves. 

We note that Wald's cosmic no-hair theorem \cite{Wald:1983ky}, which implies isotropization of the Bianchi models (except type IX model) uses SEC and DEC. Explicitly, it states that if $T_{\mu\nu}=\Lambda g_{\mu\nu}+ \tilde{T}_{\mu\nu}$ with $\Lambda>0$ and $\tilde{T}_{\mu\nu}$ satisfies SEC and DEC, then the Universe isotropizes (goes to a de Sitter space) in a few Hubble times $\sqrt{3/\Lambda}$.

\bibliographystyle{JHEP}
\bibliography{BH}

\end{document}